\documentclass[aps,prd,preprint,nofootinbib]{revtex4}

\usepackage{latexsym}
\usepackage{graphicx}
\usepackage{psfrag}
\usepackage{epsfig}
\usepackage{amsmath}

\begin{document}

\title{Perturbative Photon Fluxes Generated by High-Frequency Gravitational Waves and Their Physical
Effects}
\author{$^{1}$Fangyu Li\footnote{E-mail: cqufangyuli@hotmail.com},
$^{2}$Robert M L Baker
Jr.\footnote{E-mail:DrRobertBaker@GravWave.com}, $^{1}$Zhenyun
Fang\footnote{E-mail:zyfang@cqu.edu.cn},\\ $^{3}$Gary V.
Stephenson\footnote{E-mail:seculine@gmail.com}, $^{1}$Zhenya
Chen\footnote{E-mail:ccjaazz@yahoo.com.cn}}

\address{$^{1}$Department of Physics, Chongqing University, Chongqing 400044, P.
R. China\\ \mbox{$^{2}$GRAWAVE{\textregistered} LLC,
8123 Tuscany Avenue, Playa del Rey, California 90293, USA}\\
$^{3}$Seculine Consulting, P0 Box 925, Redondo Beach, CA 90277,
USA.}

\begin{abstract}
We consider the electromagnetic (EM) perturbative effects produced
by the high-frequency gravitational waves (HFGWs) in the GHz band in
a special EM resonance system, which consists of fractal membranes,
a Gaussian beam (GB) passing through a static magnetic field. It is
predicted, under the synchroresonance condition, coherence
modulation of the HFGWs to the preexisting transverse components of
the GB produces the transverse perturbative photon flux (PPF),which
has three novel and important properties: (1)The PPF has maximum at
a longitudinal symmetrical surface of the GB where the transverse
background photon flux (BPF) vanishes; (2) the resonant effect will
be high sensitive to the propagating directions of the HFGWs; (3)
the PPF reflected or transmitted by the fractal membrane exhibits a
very small decay compared with very large decay of the much stronger
BPF. Such properties might provide a new way to distinguish and
display the perturbative effects produced by the HFGWs. We also
discuss the high-frequency asymptotic behavior of the relic GWs in
the microwave band and the positive definite issues of their
energy-momentum pseudo-tensor .\mbox{}\\

\end{abstract}
\maketitle

\section{Introduction}

Unlike usual celestial gravitational waves (GWs) having low
frequencies, which are often a small fraction of a Hz, the relic GWs
in the microwave band ($\sim 10^8-10^{11}\mbox{Hz}$\textbf{)},
predicted by the quintessential inflationary models
(QIM)\cite{1,2,3}, the pre-big bang scenario (PBBS) and some string
cosmology scenarios \cite{4,5,6,7}, form high-frequency random
signals, their root-mean-square (rms) values of the dimensionless
amplitudes might reach up to $\sim 10^{-30}-10^{-33}/\sqrt {Hz} $,
and because of their weakness and very high-frequency properties,
they are quite different from the low-frequency GWs. The thermal
motion of plasma of stars, the interaction of the EM waves with
interstellar plasma and magnetic fields, the evaporation of
primordial black holes \cite{8}, even ultra-high-intensity lasers
\cite{9} and other high-energy laboratory schemes \cite{10,11,12}
are possible means to generate the HFGWs in the GHz band and higher
frequencies. Interaction of the HFGWs with the EM fields and the EM
detection of the HFGWs have been theoretically and experimentally
studied by many authors
\cite{13,14,15,16,17,18,19,20,21,22,23,24,25,26,27,28,29,30,31,32,33,34,35,36,37,38}.These
works include the gravitation-EM conversion in the static EM
fields(e.g., the Gertsenshtein effect and its inverse effect), the
cavity classical- and cavity quantum- electrodynamical response to
the HFGWs, resonant photon-graviton conversion, Barry's phase in the
EM detection of the HFGWs, resonant interaction of the HFGWs with
the EM wave beams, the rotation of the polarization vector of EM
wave caused by the HFGWs in the toroidal waveguide, the difference
frequency resonant response of coupled spherical cavities, etc.

Although the relic GWs have not yet been detected, we can be
reasonably sure that the Earth is bathed in the sea of these relic
GWs. Since 1978 such relic and primordial background GWs have been
of ever increasing scientific interest as many researches have shown
\cite{39,40,41,42}.

Based on high-dimensional (bulk) spacetime theories, it has also
been theoretically shown \cite{43,44}that all familiar matter fields
are constrained to live on our brane world, while gravity is free to
propagate in the extra dimensions, and the HFGWs (i.e., high-energy
gravitons) would be more capable of carrying energy from our 3-brane
world than lower-frequency GWs. It is noted that propagation of the
HFGWs may be a unique and effective way for exchanging energy and
information between two adjacent parallel brane worlds
\cite{45,46,47,48}. Moreover, if the pre-big bang scenario is
correct, then the relic GWs would be an almost unique window from
which one can look back into the universe before the big bang
\cite{6,7,49}. Although these theories and scenarios may be
controversial and whether or not they have included a fatal flaw
remains to be determined. The successful detection of the
high-frequency relic gravitational waves (HFRGWs) will certainly
shed light on many of these theories.

In this paper we shall discuss some ideas and theoretical basis for
selection and detection of the HFGWs with the predicted typical
parameters $\nu _g \sim 5\times 10^9Hz\mbox{ }(5GHz)$ and $h_{rms}
\sim {10^{-30}-10^{-33}} \mathord{\left/ {\vphantom
{{10^{-30}-10^{-33}} {\sqrt {Hz} }}} \right.
\kern-\nulldelimiterspace} {\sqrt {Hz} }$\cite{1,2,3,4,5,6,7,39,40}.
This paper includes following seven parts: (1) Introduction;
(2)Asymptotic behavior of the relic GWs in the high-frequency (the
microwave band) region and positive definite issues of the
energy-momentum pseudo-tensor of the HFRGWs; (3) The EM resonant
system to the HFRGWs, i.e., the coupling system of the fractal
membranes and a Gaussian beam (GB) passing through a static magnetic
field; (4)The EM resonant response to the HFRGWs and some numerical
estimations; (5) Selection and detection of the PPFs; (6) A very
brief review to the noise issues; (7) Concluding remarks.

\section{The high-frequency relic gravitational waves in the GHz band}

1. High-frequency asymptotic behavior of the relic GW in the
microwave band.\mbox{}\\

It is well known that each polarization component $h_{ij} (\eta
,\boldsymbol{x})$of the relic GW can be written as\cite{1, 2, 50}
\begin{equation}
\label{eq01} h_{ij} =\frac{\mu (\eta )}{a}\exp (i\boldsymbol{k}\cdot
\boldsymbol{x})e_{ij} ,
\end{equation}
The time dependent of $h$ is determined by the $\mu (\eta )$
satisfying the equation
\begin{equation}
\label{eq02} \ddot {\mu }+(k^2-{\ddot {a}} \mathord{\left/
{\vphantom {{\ddot {a}} a}} \right. \kern-\nulldelimiterspace} a)\mu
=0,
\end{equation}
where $\ddot {a}=\frac{\partial ^2a}{\partial \eta ^2},\mbox{
}a=a(\eta )$ is the cosmology scale factor, $\eta $ is the conformal
time. In fact, the Eq. (\ref{eq02}) has different exact solutions
\cite{51,52} in the different evolution stages of the Universe,
their analytic forms are often very complicated. Fortunately, for
HFRGWs in the GHz band (i.e., the relic gravitons of large
momentum), we have $k^2 \gg \left| {{\ddot {a}} \mathord{\left/
{\vphantom {{\ddot {a}} a}} \right. \kern-\nulldelimiterspace} a}
\right|$ in Eq.(\ref{eq02}), i.e., term ${\ddot {a}} \mathord{\left/
{\vphantom {{\ddot {a}} a}} \right. \kern-\nulldelimiterspace} a$
can be neglected, then the solution forms can be greatly simplified.
In this case Eq. (\ref{eq02}) has the usual periodic solution
\begin{equation}
\label{eq03} \mu (\eta )=A_1 (k)\exp (-ik\eta )+A_2 (k)\exp (ik\eta
).
\end{equation}
By using Eqs. (\ref{eq01}) and (\ref{eq03}), we have
\begin{equation}
\label{eq1} h={A_1 (k)} \mathord{\left/ {\vphantom {{A_1 (k)}
{a(\eta )}}} \right. \kern-\nulldelimiterspace} {a(\eta )}\exp
[i(\boldsymbol{k}\cdot \boldsymbol{x}-k\eta )]+{A_2 (k)}
\mathord{\left/ {\vphantom {{A_2 (k)} {a(\eta )}}} \right.
\kern-\nulldelimiterspace} {a(\eta )}\exp [i(\boldsymbol{k}\cdot
\boldsymbol{x}+k\eta )],
\end{equation}
Consequentially, the HFRGWs can be seen as the superposition of all
``monochromatic components'', Eq. (\ref{eq1}).\mbox{}\\

2. The positive definite issues of the energy-momentum pseudo-tensor
of the HFRGWs.\mbox{}\\

If the relic GWs do exist and have an observable effect, they should
have reasonable expressions for their energy-momentum pseudo-tensor
(EMPT). In particular, the energy density of the relic GWs should be
positive definite, and the momentum density components should have
reasonable physical behavior. Although the energy spectrum of the
relic GWs and their imprint on the cosmic microwave background have
been much discussed, there is little research into the complete
forms of the EMPT of the relic GWs \cite{3,53}, this research may
provide a theoretic basis for existence of relic GWs and their
detection. Unlike previous works, our attention will be only focused
on the EMPT of the HFRGWs in the GHz band, especially the positive
definite property of the energy density of them.

The relic GWs are small corrections to the background metric tensor,
the spacetime background is the de Sitter spacetime, and the metric
takes the form\cite{39,40}\mbox{}\\
\begin{equation}
\label{eq2} ds^2=a^2(\eta )[-d\eta ^2+(\delta _{ij} +h_{ij}
)dx^idx^j]=g_{\mu \nu } dx^\mu dx^\nu ,
\end{equation}
\begin{equation}
\label{eq3} g_{\mu \nu } =\bar {g}_{\mu \nu } +a^2h_{\mu \nu } ,
\end{equation}
where
\begin{equation}
\label{eq4} \bar {g}_{\mu \nu } =(-a^2,a^2,\mbox{ }a^2,\mbox{ }a^2),
\end{equation}\mbox{}\\
$\delta _{ij} $ is Kroeneker symbol. Because the sea of the HFRGWs
can be seen as the superposition of all ``monochromatic components''
of the Fourier expansion, each ``monochromatic component'', Eq.
(\ref{eq1}), contains every possible propagating direction. In this
case, we consider a single ``monochromatic wave'' propagating along
the z-axis in Cartesian coordinates without loss of generality. From
Eqs. (\ref{eq2})- (\ref{eq4}), then the metric has following form in
Cartesian coordinates\mbox{}\\
\begin{equation}
\label{eq5} g_{\mu \nu } =\left( {{\begin{array}{*{20}c}
 {-a^2}  & 0  & 0 & 0  \\
 0  & {a^2(1+h_\oplus )} & {a^2h_\otimes } & 0 \\
 0  & {a^2h_\otimes }  & {a^2(1-h_\oplus )} & 0  \\
 0  & 0  & 0  & {a^2} \\
\end{array} }} \right),
\end{equation}\mbox{}\\
From Eq. (\ref{eq5}), we have
\begin{equation}
\label{eq6}
\begin{array}{l}
 g_{00} =-a^2,\mbox{ }g_{11} =a^2\mbox{(1+}h_\oplus \mbox{), }g_{22}
=a^2\mbox{(1-}h_\oplus \mbox{), }g_{33} \mbox{=}a^2\mbox{, }g_{12}
\mbox{=}g_{21} \mbox{=}a^2h_\otimes \mbox{,} \\
 \end{array}
\end{equation}
and
\begin{equation}
\label{eq100} g=det(g_{\mu\nu})=a^8(h^2_\oplus+h^2_\otimes-1)
\end{equation}\mbox{}\\
 Expressions of the Einstein EMPT are \cite{54}
\begin{equation}
\label{eq7} \sqrt {-g} t_\mu ^\nu =\frac{c^4}{16\pi G}H_{\mu ,\sigma
}^{\nu \sigma } ,
\end{equation}
where
\begin{equation}
\label{eq8} H_\mu ^{\nu \sigma } =\frac{1}{\sqrt {-g} }g_{\mu
\lambda } [-g(g^{\nu \lambda }g^{\sigma \gamma }-g^{\sigma \lambda
}g^{\nu \gamma })]_{,\gamma } .
\end{equation}\mbox{}\\
is the super-potential. Since $h^2$ terms have been taken into
account in the determinant of metric,Eq.(\ref{eq100}),and the EMPT
for gravitational field concerns quadratic terms of $h$, the
$h^2_\oplus$ and $h^2_\otimes$ terms in the inverse of metric should
also be taken into account. By using Eqs.
(\ref{eq5}),(\ref{eq6}),(\ref{eq100}) and
$g_{\mu\alpha}g^{\alpha\nu}=\delta^\nu_\mu$, neglecting third- and
higher-orders infinitely small quantities, we obtain non-vanishing
components of $g^{\mu\nu}$ and $H^{\nu\alpha}_\mu$ in empty space as
follows
\begin{equation}
\label{eq101}
\begin{array}{l}
 g^{00} =-a^{\mbox{-}2},\mbox{ } \\ g^{11} =a^{-2}\mbox{(1-}h_\oplus+h_\oplus ^2 +h_\otimes ^2\mbox{), } g^{22}
=a^{-2}\mbox{(1+}h_\oplus+h_\oplus ^2 +h_\otimes ^2  \mbox{), }
\\ g^{33} \mbox{=}a^{-2}\mbox{, }g^{12}
\mbox{=}g^{21} \mbox{=}-a^{-2} h_\otimes \mbox{,} \\
 \end{array}
\end{equation}

\begin{eqnarray}
\label{eq9} H_0^{03}  =  - H_0^{30}  = \frac{1}{{\sqrt { - g}
}}g_{00} \left( { - gg^{00} g^{33} } \right)_{,3}  =  - 2ika^2 (h_
\oplus ^2  + h_ \otimes ^2 ),
\end{eqnarray}
\begin{eqnarray}
\label{eq10}
 H_1^{01}  &=&  - H_1^{10}  = \frac{1}{{\sqrt { - g} }}g_{11} \left( {gg^{11} g^{00} } \right)_{,0}  + \frac{1}{{\sqrt { - g} }}g_{12} \left( {gg^{12} g^{00} }
  \right)_{,0} \nonumber \\ &=& 4a\dot a - a^2 \dot h_ \oplus - 2a\dot a(h_ \oplus ^2  + h_ \otimes ^2 ) - a^2 (h_ \oplus  \dot h_ \oplus   + h_ \otimes  \dot h_ \otimes  ),
\end{eqnarray}
\begin{eqnarray}
\label{eq11}
 H_1^{02}  &=&  - H_1^{20}  = \frac{1}{{\sqrt { - g} }}g_{11} \left( {gg^{21} g^{00} } \right)_{,0}  + \frac{1}{{\sqrt { - g} }}
 g_{12} \left( {gg^{22} g^{00} } \right)_{,0}\nonumber \\ &=& a^2 (h_ \otimes  \dot h_ \oplus   - h_ \oplus  \dot h_ \otimes  ) - a^2 \dot h_ \otimes ,
\end{eqnarray}
\begin{eqnarray}
\label{eq12}
 H_1^{13}  &=&  - H_1^{31}  = \frac{1}{{\sqrt { - g} }}g_{11} \left( { - gg^{11}
 g^{33} } \right)_{,3}  + \frac{1}{{\sqrt { - g} }}g_{12} \left( { - gg^{12} g^{33} } \right)_{,3} \nonumber\\ &=&  - ika^2 (h_ \oplus   + h_ \oplus ^2  + h_ \otimes ^2 ),
\end{eqnarray}
\begin{eqnarray}
\label{eq13} H_1^{23}  =  - H_1^{32}  = \frac{1}{{\sqrt { - g}
}}g_{11} \left( { - gg^{21} g^{33} } \right)_{,3}  + \frac{1}{{\sqrt
{ - g} }}g_{12} \left( { - gg^{22} g^{33} } \right)_{,3}  =  - ika^2
h_ \otimes,
\end{eqnarray}
\begin{eqnarray}
\label{eq14} H_2^{01}  &=&  - H_2^{10}  = \frac{1}{{\sqrt { - g}
}}g_{21} \left( {gg^{11} g^{00} } \right)_{,0}  + \frac{1}{{\sqrt {
- g} }}g_{22} \left( {gg^{12} g^{00} } \right)_{,0} \nonumber\\ &=&
a^2 (h_ \oplus  \dot h_ \otimes   - \dot h_ \oplus   - \dot h_
\otimes) ,
\end{eqnarray}
\begin{eqnarray}
\label{eq15} H_2^{02}  &=&  - H_2^{20}  = \frac{1}{{\sqrt { - g}
}}g_{21} \left( {gg^{21} g^{00} } \right)_{,0}  + \frac{1}{{\sqrt {
- g} }}g_{22} \left( {gg^{22} g^{00} } \right)_{,0} \nonumber\\ &=&
4a\dot a + a^2 \dot h_ \oplus   - 2a\dot a(h_ \oplus ^2  + h_
\otimes ^2 ) - a^2 (h_ \oplus  \dot h_ \oplus   + h_ \otimes  \dot
h_ \otimes  ),
\end{eqnarray}
\begin{eqnarray}
\label{eq16} H_2^{13}  =  - H_2^{31}  = \frac{1}{{\sqrt { - g}
}}g_{21} \left( { - gg^{11} g^{33} } \right)_{,3}  + \frac{1}{{\sqrt
{ - g} }}g_{22} \left( { - gg^{12} g^{33} } \right)_{,3}  =  - ika^2
h_ \otimes,
\end{eqnarray}
\begin{eqnarray}
\label{eq17} H_2^{23}  &=&  - H_2^{32}  = \frac{1}{{\sqrt { - g}
}}g_{21} \left( { - gg^{21} g^{33} } \right)_{,3}  + \frac{1}{{\sqrt
{ - g} }}g_{22} \left( { - gg^{22} g^{33} } \right)_{,3} \nonumber\\
&=& ika^2 (h_ \oplus   - h_ \oplus ^2  - h_ \otimes ^2 ),
\end{eqnarray}
\begin{eqnarray}
H_3^{03}  &=&  - H_3^{30}  = \frac{1}{{\sqrt { - g} }}g_{33} \left(
{gg^{33} g^{00} } \right)_{,0}\nonumber \\ &=& 4a\dot a - 2a\dot
a(h_ \oplus ^2 + h_ \otimes ^2 ) - 2a^2 (h_ \oplus  \dot h_ \oplus +
h_ \otimes \dot h_ \otimes) ,
\end{eqnarray}
\mbox{}\\ From Eqs. (\ref{eq1})-(\ref{eq100}),
(\ref{eq101})and(\ref{eq9}), we obtain the energy
density and energy flux density as follows, respectively \mbox{}\\
\begin{equation}
\label{eq18}
t_0^0 =\frac{c^4}{16\pi G\sqrt {-g} }H_{0,\sigma }^{0\sigma }
=\frac{c^4k^2}{4\pi Ga^2}(h_\oplus ^2 +h_\otimes ^2 ),
\end{equation}\mbox{}\\
\begin{equation}
\label{eq19}
ct_0^1 =cH_{0,\sigma }^{1\sigma } =ct_0^2 =cH_{0,\sigma }^{2\sigma } =0,
\end{equation}\mbox{}\\
\begin{equation}
\label{eq20}
ct_0^3 =\frac{c^5}{16\pi G\sqrt {-g} }H_{0,\sigma }^{3\sigma }
=\frac{ikc^5}{4\pi Ga^3}[\dot {a}(h_\oplus ^2 +h_\otimes ^2 )+a(h_\oplus
\dot {h}_\oplus +h_\otimes \dot {h}_\otimes )].
\end{equation}\mbox{}\\
For the ``monochromatic components'' of propagating along the x- and
y-axises, we have the same expression for the energy density, Eq.
(\ref{eq18}), but $t_0^2 =t_0^3 =0$ and $t_0^1 =t_0^3
=0$,respectively. The energy flux density also has the same form,
Eq. (\ref{eq20}). Thus, the energy density of the HFRGWs is positive
definite, and the energy flux densities have reasonable physical
behavior in the conformal time coordinates. If we integrate the EMPT
for all ``the monochromatic components'' of the HFRGWs, then we can
find that the EMPT is homogeneous and isotropic. Riazuelo and Uzan
\cite{3}obtained an expression for the EMPT of the relic GWs in the
momentum space, the average values of such expressions for the EMPT
have reasonable physical behavior. However, Eq. (\ref{eq1}) shows
that the stochastic relic GWs background contains every possible
propagating direction, and because of stochastic fluctuation of the
amplitudes of the HFRGWs over their bandwidth, detection of the
HFRGWs will be more difficult than that of the monochromatic plane
GWs. In this case, can the HFRGWs be selected and measured? In
particular, if two HFRGWs have the same amplitude and frequency, but
propagate along the exactly opposite directions (standing wave),
will their effect be canceled and nullified? We shall show that in
our EM system the EM perturbation produced by the HFRGWs, which
propagate along the positive and negative directions of the
symmetrical axis (the $z$-axis ) of the GB, will be non-symmetric
and the physical effect generated by the HFRGWs propagating along
other directions will be also quite different, even if they satisfy
the resonant condition ($\omega _e =\omega _g )$, and only the HFRGW
component propagating along the positive direction of the
symmetrical $z-$axis of the GB can generate an optimal resonant
response. Thus our EM system design will be very sensitive to the
propagating directions as well as the frequencies of the HFRGWs, and
it may provide a HFRGW map of the celestial sphere (similar to the
map of the relic microwave background provided by the Wilkinson
Microwave Anisotropic Probe or WMAP).

\section{ The electromagnetic resonant system: the coupling system of
the fractal membranes and the Gaussian beam (GB) passing though a
static magnetic field}

Our EM system consists of the GB of a fundamental frequency mode
\cite{55} operating in the GHz band immersed in a static magnetic
field, with a new-type of fractal membranes \cite{56,57,58} to focus
PPF signal along the detection axis. In order to consider resonant
response to the HFRGWs in the laboratory frame of reference, all
parameters of the EM system should be values in the frame of
reference. The general form of the GB of a fundamental frequency
mode [55] is\mbox{}\\
\begin{equation}
\label{eq21}
\psi =\frac{\psi _0 }{\sqrt {1+(z/f)^2} }\exp (-\frac{r^2}{W^2})\exp \left\{
{i[(k_e z-\omega _e t)-\tan ^{-1}\frac{z}{f}+\frac{k_e r^2}{2R}+\delta ]}
\right\},
\end{equation}\mbox{}\\
where $r^2=x^2+y^2$, $k_e =2\pi /\lambda _e $, $f=\pi W_0^2 /\lambda
_e $, $W=W_0 [1+(z/f)^2]^{1/2}$, $R=z+f^2/z$, $\psi _0 $ is the
amplitude of electric (or magnetic) field of the GB, $W_{^0} $ is
the minimum spot radius, $R$ is the curvature radius of the wave
fronts of the GB at z, $\omega _e $ is the angular frequency,
$\lambda _{e}$ is the EM wavelength, the $z$-axis is the symmetrical
axis of the GB, and $\delta $ is an arbitrary phase factor. For the
resonant response to a HFRGW, $\delta $ is the phase difference
between the GB and the resonant component of the HFRGW. Using a new
approach, different from Refs. [21, 22], we choose the GB with the
double transverse polarized electric modes (DTEM), and utilize the
coupling effect between the fractal membrane in the GHz band and the
GB passing through a static magnetic field. Indeed, the GBs with the
DTEM exhibit more realizable modes, they have been extensively
discussed and applied in the closed resonant cavities, open
resonators and free space \cite{55,59,60,61}, including the
standing-wave-type and traveling-wave-type GBs. Moreover, a very
important property of the EM system is that the PPF (signal)
reflected or transmitted by the fractal membranes exhibits a very
small decay \cite{56,57,58} in transit to the detectors
(high-sensitivity microwave photon flux receivers) compared with the
very large decay (typical Gaussian decay rate) of the much stronger
BPF. This property provides a possibility to distinguish them in
some suitable regions.

If the static magnetic field pointing along the y-axis is localized in the
region $-l_1 \le z\le l_2 $, setting $\tilde {E}_x^{(0)} =\psi =\psi _x $
and using divergenceless condition $\nabla \cdot {\rm {\bf
E}}=\frac{\partial \psi _x }{\partial x}+\frac{\partial \psi _y }{\partial
y}=0$ and ${\rm {\bf \tilde {B}}}^{(0)}=-\frac{i}{\omega _e }\nabla \times
{\rm {\bf \tilde {E}}}^{(0)}$ (we use MKS units), then we have\mbox{}\\
\begin{equation}
\label{eq22}
\tilde {E}_x^{(0)} =\psi =\psi _x ,
\quad
\tilde {E}_y^{(0)} =\psi _y =-\int {\frac{\partial \psi _x }{\partial x}}
dy=2x(\frac{1}{W^2}-i\frac{k_e }{2R})\int {\psi _x } dy,
\quad
\tilde {E}_z^{(0)} =0,
\end{equation}
\begin{equation}
\label{eq23}
\tilde {B}_x^{(0)} =\frac{i}{\omega _e }\frac{\partial \psi _y }{\partial
z},
\quad
\tilde {B}_y^{(0)} =-\frac{i}{\omega _e }\frac{\partial \psi _x }{\partial
z},
\quad
\tilde {B}_z^{(0)} =\frac{i}{\omega _e }(\frac{\partial \psi _x }{\partial
y}-\frac{\partial \psi _y }{\partial x}),
\end{equation}
and
\begin{equation}
\label{eq24}
\hat {B}^{(0)}=\left\{ {\begin{array}{l}
 \hat {B}_y^{(0)} \mbox{ (}-l_1 \le z\le l_2 \mbox{) ,} \\
 0\mbox{ (}z\le -l_1 \mbox{ and }z>l_2 ), \\
 \end{array}} \right.
\end{equation}\mbox{}\\
where the superscript 0 denotes the background EM fields, the
notations $\sim $ and $\wedge$ stand the time-dependent and static
EM fields, respectively. For the high-frequency EM power flux (or in
quantum language: photon flux), only non-vanishing average values of
this with respect to time have an observable effect. From
Eqs. (\ref{eq21}), (\ref{eq22}), and (\ref{eq23}), one finds\mbox{}\\
\begin{equation}
\label{eq25}
n_x^{(0)} =\frac{1}{\hbar \omega _e }\langle \frac{1}{\mu _0 }(\tilde
{E}_y^{(0)} \tilde {B}_z^{(0)} )\rangle =\frac{1}{2\mu _0 \hbar \omega _e
}Re\left\{ {\psi _y^\ast [\frac{i}{\omega _e }(\frac{\partial \psi _x
}{\partial y}-\frac{\partial \psi _y }{\partial x})]} \right\}=f_x
^{(0)}\exp (-\frac{2r^2}{W^2}),
\end{equation}\mbox{}\\
\begin{equation}
\label{eq26}
n_y^{(0)} =-\frac{1}{\hbar \omega _e }\langle \frac{1}{\mu _0 }(\tilde
{E}_x^{(0)} \tilde {B}_z^{(0)} )\rangle =\frac{1}{2\mu _0 \hbar \omega _e
}Re\left\{ {\psi _x^\ast [\frac{i}{\omega _e }(\frac{\partial \psi _y
}{\partial x}-\frac{\partial \psi _x }{\partial y})]} \right\}=f_y^{(0)}
\exp (-\frac{2r^2}{W^2}),
\end{equation}\mbox{}\\
\begin{eqnarray}
\label{eq27}
n_z^{(0)} &=&\frac{1}{\hbar \omega _e }\langle
\frac{1}{\mu _0 }(\tilde {E}_x^{(0)} \tilde {B}_y^{(0)}
)-\frac{1}{\mu _0 }(\tilde
{E}_y^{(0)} \tilde {B}_x^{(0)} )\rangle\nonumber\\
&=&\frac{1}{2\mu _0 \hbar \omega _e }Re\left\{ {\psi _x^\ast
[\frac{i}{\omega _e }(\frac{\partial \psi _y }{\partial z})]+\psi
_y^\ast [\frac{i}{\omega _e }(\frac{\partial \psi _x }{\partial
z})]} \right\}=f_z^{(0)} \exp (-\frac{2r^2}{W^2}),
\end{eqnarray}\mbox{}\\
where $\hbar \omega _e $ is the energy of single photon, $n_x^{(0)}
$, $n_y^{(0)} $ and $n_z^{(0)} $ represent the average values of the
$x$-, $y- $and $z$- components of the BPF densities, in units of
photons per second per square meter, propagating along the x-, y-
and z-axes, respectively, the angular brackets denote the average
over time, $f_x^{(0)} $, $f_y^{(0)} $ and $f_z^{(0)} $ are the
functions of $\psi _0 $, $W_0 $, $\omega _e $, $r$ and $z$. Because
of the non-vanishing $n_x^{(0)} $ and $n_y^{(0)} $, the GB will be
asymptotically spread as $\vert z\vert $ increases (i.e., the
irradiance surface of the GB spreads out in the + z and -- z
directions).

\section{The EM resonant response to the HFRGWs}

For the EM resonant response in the laboratory frame of reference, we should
use the intervals of laboratory time (i.e., $cdt=a(\eta )d\eta )$ and
laboratory frequency of the HFRGWs. In this case, Eq.(\ref{eq1}) can be written as
\begin{equation}
\label{eq28} h({\bf x},t) = {{A(k_g )} \mathord{\left/
 {\vphantom {{A(k_g )} {a(t)}}} \right.
 \kern-\nulldelimiterspace} {a(t)}}\exp [i({\bf k}_g {\rm{.}}{\bf x}{\rm{ - }}\omega _g t)] + {{B(k_g )} \mathord{\left/
 {\vphantom {{B(k_g )} {a(t)}}} \right.
 \kern-\nulldelimiterspace} {a(t)}}\exp [i({\bf k}_g {\rm{.}}{\bf x}{\rm{ + }}\omega _g t)
,
\end{equation}
where $A \mathord{\left/ {\vphantom {A a}} \right.
\kern-\nulldelimiterspace} a$ and $B \mathord{\left/ {\vphantom {B
a}} \right. \kern-\nulldelimiterspace} a$ are the stochastic values
of the amplitudes of the HFRGWs in the laboratory frame of
reference, $k_g $ and $\omega _g $ are the corresponding wave vector
and angular frequency in the frame of reference. Eq. (\ref{eq28})
can be seen as the approximate form of each ``monochromatic
polarization component'' of the HFRGWs in the GHz band. In our EM
system, since only ``monochromatic component'' of the HFGW
propagating along the positive direction of the symmetrical axis
(the z-axis) of the GB generates an optimal resonant response (see
Section \uppercase\expandafter{\romannumeral5}), our attention will
be focused to a circular polarized ``monochromatic
component'' of the HFRGW in the z-direction, i.e., \mbox{} \\
\begin{equation}
\label{eq102}  \begin{array}{l}
 h_ \oplus   = h_{11}  =  - h_{22}  = A_ \oplus  \exp \left[ {i\left( {k_g z - \omega _g t} \right)} \right], \\
 h_ \otimes   = h_{12}  = h_{21}  = iA_ \otimes  \exp \left[ {i\left( {k_g z - \omega _g t} \right)} \right], \\
 \end{array} \
 \end{equation} \mbox{}\\
 where $A_ \oplus  ,A_ \otimes   \approx {{A\left( {k_g } \right)} \mathord{\left/
 {\vphantom {{A\left( {k_g } \right)} {a\left( t \right)}}} \right.
 \kern-\nulldelimiterspace} {a\left( t \right)}}$ [see
 Eq.(\ref{eq28})].Using the electrodynamical equations in the curved
 spacetime \mbox{} \\
\begin{equation}
\label{eq29}
\frac{1}{\sqrt {-g} }\frac{\partial }{\partial x^\nu }(\sqrt {-g} g^{\mu
\alpha }g^{\nu \beta }F_{\alpha \beta } )=\mu _0 J^\mu ,
\end{equation}
\begin{equation}
\label{eq30}
\nabla _\alpha F_{\mu \nu } +\nabla _\nu F_{\alpha \mu } +\nabla _\mu F_{\nu
\alpha } =0,
\end{equation}\mbox{} \\
we can describe the EM perturbation produced by the HFRGWs in the EM
system, where $F_{\mu \nu } $ is the EM field tensor, and $F_{\mu
\nu } =F_{\mu \nu }^{(0)} +\tilde {F}_{\mu \nu }^{(1)} $, $F_{\mu
\nu }^{(0)} $ and $\tilde {F}_{\mu \nu }^{(1)} $ represent the
background and first-order perturbative EM fields respectively in
the presence of the HFRGWs. $J^\mu $ indicates the four-dimensional
electric current density. For the EM response in vacuum, $J^\mu =0$
in Eq. (\ref{eq29}). Because of the weak field property of the
HFRGWs, the perturbation methods will still be valid. Using
Eqs.(\ref{eq6}),(\ref{eq100}) and (\ref{eq101}), then
Eqs.(\ref{eq29}), (\ref{eq30}) can be reduced to \mbox{} \\

\begin{equation}
\label{eq103}\frac{\partial }{{\partial x^\nu  }}\left[ {a^4 g^{\mu
\alpha } g^{\nu \beta } \left( {F_{\alpha \beta }^{(0)}  + \tilde
F_{\alpha \beta }^{(1)} } \right)} \right] = \frac{{\left( {h_
\oplus  \frac{{\partial h_ \oplus  }}{{\partial x_\nu  }} + h_
\otimes  \frac{{\partial h_ \otimes  }}{{\partial x_\nu  }}}
\right)\left[ {a^4 g^{\mu \alpha } g^{\nu \beta } \left( {F_{\alpha
\beta }^{(0)}  + \tilde F_{\alpha \beta }^{(1)} } \right)}
\right]}}{{1 - h_ \oplus ^2  - h_ \otimes ^2 }},
\end{equation}

\begin{equation}
\label{eq104} \nabla _\alpha  \left( {F_{\mu \nu }^{(0)}  + \tilde
F_{\mu \nu }^{(1)} } \right) + \nabla _\nu  \left( {F_{\alpha \mu
}^{(0)}  + \tilde F_{\alpha \mu }^{(1)} } \right) + \nabla _\mu
\left( {F_{\nu \alpha }^{(0)}  + \tilde F_{\nu \alpha }^{(1)} }
\right) = 0,
\end{equation} \mbox{} \\ Unlike plane monochromatic GWs, the amplitudes of the relic
GW in Eq. (\ref{eq28}) are not constant, in this case solving
Eqs.(\ref{eq103}) and (\ref{eq104}) will often be difficult. In our
case, fortunately, since this is the EM response in the GHz band,
and considering Eq.(\ref{eq102}), the following equivalent relations
would be valid provided $\omega _g \gg \dot {a}/a$, i.e.,
\begin{equation}
\label{eq31}
\frac{\partial }{\partial t}\to \mp i\omega _g ,
\quad
\nabla \to i{\rm {\bf k}}_g .
\end{equation}
In this case the process of solving Eqs.(\ref{eq103}) and
(\ref{eq104}) can be greatly simplified without excluding their
essential physical features.

Introducing Eqs.(\ref{eq6}),(\ref{eq100}),(\ref{eq101}) and
(\ref{eq102}) into Eqs.(\ref{eq103}) and (\ref{eq104}), considering
$\left| {h_ \oplus  } \right|,\left| {h_ \otimes  } \right| \ll 1$,
using the equivalent relations, Eq.(\ref{eq31}) and neglecting
high-order infinitely small quantities, the first-order perturbative
EM fields generated by the direct interaction of the z-component of
a certain ``monochromatic wave,'' Eq.(\ref{eq102}), with the static
magnetic field $\hat {B}_y^{(0)} $ can be given by \cite{14,21,28}
\mbox{} \\
\begin{equation}
\label{eq32}
\begin{array}{l}
 \tilde {E}_x^{(1)} =\frac{i}{2}A_\oplus \hat {B}_y^{(0)} k_g c(z+l_1 )\exp
[i(k_g z-\omega _g t)]+\frac{1}{4}A_\oplus \hat {B}_y^{(0)} c\exp
[i(k_g
z+\omega _g t)], \\
 \tilde {B}_y^{(1)} =\frac{i}{2}A_\oplus \hat {B}_y^{(0)} k_g (z+l_1 )\exp
[i(k_g z-\omega _g t)]-\frac{1}{4}A_\oplus \hat {B}_y^{(0)} \exp
[i(k_g
z+\omega _g t)], \\
 \tilde {E}_y^{(1)} =-\frac{1}{2}A_\otimes \hat {B}_y^{(0)} k_g c(z+l_1
)\exp [i(k_g z-\omega _g t)]+\frac{i}{4}A_\otimes \hat {B}_y^{(0)}
c\exp
[i(k_g z+\omega _g t)], \\
 \tilde {B}_x^{(1)} =\frac{1}{2}A_\otimes \hat {B}_y^{(0)} k_g (z+l_1 )\exp
[i(k_g z-\omega _g t)]+\frac{i}{4}A_\otimes \hat {B}_y^{(0)} \exp
[i(k_g
z+\omega _g t)], \\
 \end{array}
\end{equation}\mbox{} \\
where $A_\oplus ,A_\otimes \approx A(k_g )/a(t)$, [see, Eq.
(\ref{eq28})],$-l_1 \le z\le l_2 $. Equation (\ref{eq32}) shows that
the first-order perturbative EM fields have a space accumulation
effect ($\propto z)$ in the interacting region: this is because the
GWs (gravitons) and EM waves (photons) have the same propagating
velocity, so that the two waves can generate an optimum coherent
effect in the propagating direction \cite{14, 28}. Such results and
the calculation by Feynman perturbation techniques in Ref.\cite{14}
are self-consistent. In our EM system, we shall neglect the EM
perturbation solution which describes the EM perturbation
propagating along the negative direction of the z=axis since it can
not satisfy the boundary condition $\left. {\tilde F_{\mu \nu
}^{(1)} } \right|_{z = - l_1 }=0$. Obviously, this is typical
inverse Gertsenshtein effect \cite{13}. From Eqs.
(\ref{eq22}),(\ref{eq23}), (\ref{eq24}) and (\ref{eq32}), the total
EM field tensors in the
presence of the HFRGW can be written as\mbox{} \\

$F_{\mu \nu } =F_{\mu \nu }^{(0)} +\tilde {F}_{\mu \nu }^{(1)}$
\begin{equation}
\label{eq33}
=\left( {{\begin{array}{*{20}c}
 0 & {\frac{1}{c}(\tilde {E}_x^{(0)} +\tilde {E}_x^{(1)} )}&
{\frac{1}{c}(\tilde {E}_y^{(0)} +\tilde {E}_y^{(1)} )}& 0  \\
 {-\frac{1}{c}(\tilde {E}_x^{(0)} +\tilde {E}_x^{(1)} )}& 0 &
{-\tilde {B}_z^{(0)} }& {\hat {B}_y^{(0)} +\tilde {B}_y^{(0)}
+\tilde {B}_y^{(1)} }\\
 {-\frac{1}{c}(\tilde {E}_y^{(0)} +\tilde {E}_y^{(1)} )}& {\tilde
{B}_z^{(0)} }& 0 & {-(\tilde {B}_x^{(0)} +\tilde {B}_x^{(1)}
)}\\
 0& {-(\hat {B}_y^{(0)} +\tilde {B}_y^{(0)} +\tilde {B}_y^{(1)} )}
& {\tilde {B}_x^{(0)} +\tilde {B}_x^{(1)} }& 0 \\
\end{array} }} \right).
\end{equation}\mbox{} \\
In our exemplar EM system we have chosen the GB power of $P=10W$ and
the static magnetic field of $\hat {B}_y^{(0)} =3T$, then
corresponding magnetic field amplitude of the GB is only $\tilde
{B}^{(0)}\sim 10^{-5}T$, so the ratio of $\tilde {B}^{(0)}$ and the
background static magnetic field $\hat {B}_y^{(0)} $ is roughly
$\tilde {B}^{(0)}/\hat {B}_y^{(0)} \sim 10^{-5}$. In this case we
have neglected the perturbation EM fields produced by the directed
interaction of the HFRGW with the GB.

Using the generic expression of the energy-momentum tensor of the EM fields
in GW fields
\begin{equation}
\label{eq34}
T^{\mu \nu }=\frac{1}{\mu _0 }(-F_\alpha ^\mu F^{\nu \alpha
}+\frac{1}{4}g^{\mu \nu }F_{\alpha \beta } F^{\alpha \beta }),
\end{equation}\mbox{} \\
we can calculate the perturbation to the energy-momentum of the EM
fields in the GW fields. Because of the weak field property of the
HFRGWs, the energy-momentum tensor $T^{\mu \nu } $ can also be
decomposed into
\begin{equation}
\label{eq35}
T^{\mu \nu }=\mathop {T^{\mu \nu }}\limits^{(0)} +\mathop {T^{\mu \nu
}}\limits^{(1)} +\mathop {T^{\mu \nu }}\limits^{(2)} ,
\end{equation}
where $\mathop {T^{\mu \nu }}\limits^{(0)} $ is the energy-momentum
tensor of the background EM fields, $\mathop {T^{\mu \nu
}}\limits^{(1)} $and $\mathop {T^{\mu \nu }}\limits^{(2)} $ are
first- and second- order perturbations to $\mathop {T^{\mu \nu
}}\limits^{(0)} $ in the presence of the HFRGW. From Eqs.
(\ref{eq34}) and (\ref{eq35}), $\mathop {T^{\mu \nu }}\limits^{(0)}
$,$\mathop {T^{\mu \nu }}\limits^{(1)} $ and $\mathop {T^{\mu \nu
}}\limits^{(2)} $ can be written as\mbox{} \\
\begin{equation}
\label{eq36}
\mathop {T^{\mu \nu }}\limits^{(0)} =\frac{1}{\mu _0 }[-F_\alpha ^{\mu (0)}
F^{\nu \alpha (0)}+\frac{1}{4}\delta ^{\mu \nu }F_{\alpha \beta }^{(0)}
F^{\alpha \beta (0)}],
\end{equation}
\begin{eqnarray}
\label{eq37} \mathop {T^{\mu \nu }}\limits^{(1)} =&\frac{1}{ \mu _0
}&[-(F_\alpha ^{\mu (0)} \tilde {F}^{\nu \alpha (1)}+\tilde {F}^{\mu
{(1)}} _\alpha \mbox{ }F^{\nu \alpha (0)})+\frac{1}{4}\delta ^{\mu
\nu }(\tilde {F}_{\alpha \beta }^{(1)} F^{\alpha \beta
(0)}+F_{\alpha \beta }^{(0)} \tilde {F}^{\alpha
\beta (1)}) \\
&-&\frac{1}{4}h^{\mu \nu }F_{\alpha \beta }^{(0)} F^{\alpha \beta
(0)}],\nonumber
\end{eqnarray}
\begin{equation}
\label{eq38} \mathop {T^{\mu \nu }}\limits^{(2)} =\frac{1}{\mu _0
}[-(\tilde {F}_\alpha ^{\mu (1)} \tilde {F}^{\nu \alpha
(1)}+\frac{1}{4}\delta ^{\mu \nu }\tilde {F}_{\alpha \beta }^{(1)}
\tilde {F}^{\alpha \beta (1)}-\frac{1}{4}h^{\mu \nu }(F_{\alpha
\beta }^{(0)} \tilde {F}^{\alpha \beta (1)}+\tilde {F}_{\alpha \beta
}^{(1)} F^{\alpha \beta (0)})],
\end{equation}\mbox{} \\
Eqs. (\ref{eq32}), (\ref{eq33}), (\ref{eq37}), and (\ref{eq38}) show
that the first-order perturbation $\tilde {F}_{\mu \nu }^{(1)} $ of
the EM fields tensor contains only the first-order term of the
metric h, thus $\mathop {T^{\mu \nu }}\limits^{(1)} $ is
proportional to the first-order terms of h, while $\mathop {T^{\mu
\nu }}\limits^{(2)} $is proportional to the second-order terms of h.
Because the expected amplitude of HFRGWs in the GHz band would be
only h$\sim$10$^{-28}$-10$^{-33}$/$\sqrt {Hz} $
\cite{1,2,7,8,40,52},then for nonvanishing $\mathop {T^{\mu \nu
}}\limits^{(0)} $, $\mathop {T^{\mu \nu }}\limits^{(1)} $and
$\mathop {T^{\mu \nu }}\limits^{(2)} $, we have\mbox{} \\
\begin{equation}
\label{eq39}
\vert \mathop {T^{\mu \nu }}\limits^{(0)} \vert \gg \vert \mathop {T^{\mu
\nu }}\limits^{(1)} \vert \gg \vert \mathop {T^{\mu \nu }}\limits^{(2)}
\vert .
\end{equation}\mbox{} \\
In this case, for the effect of the HFRGW, we are interested in
$\mathop {T^{\mu \nu }}\limits^{(1)} $ but not in $\mathop {T^{\mu
\nu }}\limits^{(0)} $ and $\mathop {T^{\mu \nu }}\limits^{(2)} $.
Considering the transverse and traceless (TT) gauge condition
($h^{11}=-h^{22}=h_\oplus ,\mbox{ }h^{12}=h^{21}=h_\otimes ,h_i^i
=0, h^{01}=h^{02}=h^{03}=h^{13}=h^{23}=h^{33}=0)$, all novanishing
components of the first-order perturbation to $T^{\mu \nu }
$generated by a ``monochromatic component'' propagating along the
z-axis of the HFRGW can be written as
\begin{equation}
\label{eq40}
\mathop {T^{00}}\limits^{(1)} =\frac{1}{\mu _0 }[-(F_\alpha ^{0(0)} \tilde
{F}^{0\alpha (1)}+\tilde {F}_\alpha ^{0(1)} F^{0\alpha
(0)})+\frac{1}{4}(\tilde {F}_{\alpha \beta }^{(1)} F^{\alpha \beta
(0)}+F_{\alpha \beta }^{(0)} \tilde {F}^{\alpha \beta (1)})],
\end{equation}
\begin{equation}
\label{eq41}
\mathop {T^{01}}\limits^{(1)} =-\frac{1}{\mu _0 }(F_\alpha ^{0(0)} \tilde
{F}^{1\alpha (1)}+\tilde {F}_\alpha ^{0(1)} F^{1\alpha (0)}),
\end{equation}
\begin{equation}
\label{eq42} \mathop {T^{02} }\limits^{(1)}  =  - \frac{1}{{\mu _0
}}(F_\alpha ^{0(0)} \tilde F^{2\alpha (1)}  + \tilde F_\alpha
^{0(1)} F^{2\alpha (0)} ),
\end{equation}
\begin{equation}
\label{eq43} \mathop {T^{03}}\limits^{(1)} =-\frac{1}{\mu _0
}(F_\alpha ^{0(0)} \tilde {F}^{3\alpha (1)}+\tilde F_\alpha ^{0(1)}
F^{3\alpha (0)}),
\end{equation}
\begin{equation}
\label{eq44}
\mathop {T^{11}}\limits^{(1)} =\frac{1}{\mu _0 }[-(F_\alpha ^{1(0)} \tilde
{F}^{1\alpha (1)}+\tilde {F}_\alpha ^{1(1)} F^{1\alpha
(0)})+\frac{1}{4}(\tilde {F}_{\alpha \beta }^{(1)} F^{\alpha \beta
(0)}+F_{\alpha \beta }^{(0)} \tilde {F}^{\alpha \beta
(1)})-\frac{1}{4}h^{11}F_{\alpha \beta }^{(0)} F^{\alpha \beta (0)}],
\end{equation}
\begin{equation}
\label{eq45}
\mathop {T^{22}}\limits^{(1)} =\frac{1}{\mu _0 }[-(F_\alpha ^{2(0)} \tilde
{F}^{2\alpha (1)}+\tilde {F}_\alpha ^{2(1)} F^{2\alpha
(0)})+\frac{1}{4}(\tilde {F}_{\alpha \beta }^{(1)} F^{\alpha \beta
(0)}+F_{\alpha \beta }^{(0)} \tilde {F}^{\alpha \beta
(1)})-\frac{1}{4}h^{22}F_{\alpha \beta }^{(0)} F^{\alpha \beta (0)}],
\end{equation}
\begin{equation}
\label{eq46}
\mathop {T^{33}}\limits^{(1)} =\frac{1}{\mu _0 }[-(F_\alpha ^{3(0)} \tilde
{F}^{3\alpha (1)}+\tilde {F}_\alpha ^{3(1)} F^{3\alpha
(0)})+\frac{1}{4}(\tilde {F}_{\alpha \beta }^{(1)} F^{\alpha \beta
(0)}+F_{\alpha \beta }^{(0)} \tilde {F}^{\alpha \beta (1)})],
\end{equation}
\begin{equation}
\label{eq47}
\mathop {T^{12}}\limits^{(1)} =\mathop {T^{21}=}\limits^{(1)} -\frac{1}{\mu
_0 }[(F_\alpha ^{1(0)} \tilde {F}^{2\alpha (1)}+\tilde {F}_\alpha ^{1(1)}
F^{2\alpha (0)})+\frac{1}{4}h^{12}F_{\alpha \beta }^{(0)} F^{\alpha \beta
(0)}],
\end{equation}
\begin{equation}
\label{eq48}
\mathop {T^{13}}\limits^{(1)} =\mathop {T^{31}=}\limits^{(1)} -\frac{1}{\mu
_0 }[F_\alpha ^{1(0)} \tilde {F}^{3\alpha (1)}+\tilde {F}_\alpha ^{1(1)}
F^{3\alpha (0)}],
\end{equation}
\begin{equation}
\label{eq49} \mathop {T^{23} }\limits^{(1)}  = \mathop {T^{32}  =
}\limits^{(1)}  - \frac{1}{{\mu _0 }}(F_\alpha ^{2(0)} \tilde
F^{3\alpha (1)}  + \tilde F_\alpha ^{2(1)} F^{3\alpha (0)} ),
\end{equation}\mbox{} \\
where $\mathop {T^{00}}\limits^{(1)} $expresses the first-order
perturbation to the energy density of the EM fields, $\mathop
{T^{01}}\limits^{(1)} $, $\mathop {T^{02}}\limits^{(2)} $ and
$\mathop {T^{03}}\limits^{(3)} $ indicate the first-order
perturbations to the power flux densities of the EM fields in the
x-, y- and z- directions, respectively, while $\mathop
{T^{11}}\limits^{(1)} $, $\mathop {T^{22}}\limits^{(1)} $, $\mathop
{T^{33}}\limits^{(1)} $, $\mathop {T^{12}}\limits^{(1)} $, $\mathop
{T^{13}}\limits^{(1)} $and $\mathop {T^{23}}\limits^{(1)} $
represent the first-order perturbations to the momentum flux density
components of the EM fields.

By using Eqs. (\ref{eq21})-(\ref{eq24}) and
(\ref{eq32})-(\ref{eq34}), we can calculate the first-order PPFs
produced by the HFRGW. We shall focus our attention to the
01-component $\mathop {T^{01}}\limits^{(1)} $[see Eq. (\ref{eq41})]
of the first-order perturbation: it expresses the x-component of the
power flux density (Poynting vector) of the EM fields, i.e., the
first-order perturbative power flux density generated by the
coherent modulation of the preexisting $x$-component of background
power flux. Thus, the corresponding first-order PPF will be $c/\hbar
\omega _e \mathop {T^{01}}\limits^{(1)} $. In this case, although we
do not know value of the initial phase of ``the resonant
monochromatic component'' of the HFRGW in the laboratory frame of
reference due to its random distribution, setting the phase
difference $\delta =\pi $/2 will always be possible by regulating
the phase of the GB. The x-component of PPF generated by the
coherent synchro-resonance ($\omega _e =\omega _g )$ between the
perturbative EM fields, Eq. (\ref{eq32}) and the GB, Eqs.
(\ref{eq21})- (\ref{eq23}), can then
be expressed in the following form\mbox{} \\
\begin{eqnarray}
\label{eq50} n_x^{(1)}&=& \frac{c}{{\hbar \omega _e }}<{\mathop
{T^{01} }\limits^{(1)} }> _{\omega _e  = \omega _g } = -
\frac{c}{{\mu _0 \hbar \omega _e }}\left\langle{F_\alpha ^{0(0)}
\tilde F^{1\alpha (1)}  + \tilde F_\alpha ^{0(1)} F^{1\alpha
(0)}}\right\rangle _{\omega
_e  = \omega _g } \nonumber \mbox{} \\ \nonumber\\
&=&\frac{1}{\hbar \omega _e }\langle \frac{1}{\mu _0 }\tilde
{E}_y^{(1)} \tilde {B}_z^{(0)} \rangle _{\omega _e =\omega _g }
=\frac{1}{2\mu _0 \hbar \omega _e }Re\left\{ {\tilde {E}_y^{(1)\ast
} \left[ {\frac{i}{\omega _e }(\frac{\partial \psi _x }{\partial
y}-\frac{\partial \psi _y }{\partial x})} \right]} \right\}_{\omega
_e =\omega _g } \nonumber \mbox{} \\ \nonumber\\
&=&-\frac{1}{{\hbar \omega _e }} \cdot \left\{ {\frac{{A_ \otimes  \hat B_y^{(0)} \psi _0 k_g y(z + l_1 )}}{{4\mu _0 [1 + (z/f)^2 ]^{1/2} (z + f^2 /z)}}} \right.\sin \left( {\frac{{k_g r^2 }}{{2R}} - \tan ^{ - 1} \frac{z}{f}} \right) + \frac{{A_ \otimes  \hat B_y^{(0)} \psi _0 y(z + l_1 )}}{{2\mu _0 W_0^2 [1 + (z/f)^2 ]^{3/2} }}  \nonumber\mbox{} \\ \nonumber \\
&&\left. {\cos \left( {\frac{{k_g r^2 }}{{2R}} - \tan ^{ - 1} \frac{z}{f}} \right)} \right\}\exp \left( { - \frac{{r^2 }}{{W^2 }}} \right) - \frac{1}{{\hbar \omega _e }}\left\{ {\left( {1 - \frac{{4x^2 }}{{W^2 }}} \right)\frac{{A_ \otimes  \hat B_y^{(0)} \psi _0 k_g (z + l_1 )}}{{4\mu _0 R{\rm{ }}[1 + (z/f)^2 ]^{1/2} }}} \right. \cdot  \nonumber\mbox{} \\ \nonumber\\
&&\left[ {F_1 (y)\sin \left( {\frac{{k_g x^2 }}{{2R}} - \tan ^{ - 1}
\frac{z}{f}} \right) + F_2 (y)\cos \left( {\frac{{k_g x^2 }}{{2R}} -
\tan ^{ - 1} \frac{z}{f}} \right)} \right] + \left[ {\frac{2}{{W^2
}} + } \right. \nonumber\mbox{} \\\nonumber &&\left. {\left(
{\frac{{k_g ^2 }}{{R^2 }} - \frac{4}{{W^4 }}} \right)x^2 } \right]
\cdot \frac{{A_ \otimes  \hat B_y^{(0)} \psi _0 (z + l_1 )}}{{4\mu
_0 [1 + (z/f)^2 ]^{1/2} }}\left[ {F_1 (y)\cos \left( {\frac{{k_g x^2
}}{{2R}} - \tan ^{ - 1} \frac{z}{f}} \right)} \right.
\nonumber\\\mbox{} \\\nonumber &&- F_2 (y)\left. {\left. {\sin
\left( {\frac{{k_g x^2 }}{{2R}} - \tan ^{ - 1} \frac{z}{f}} \right)}
\right]} \right\}\exp \left( { - \frac{{x^2 }}{{W^2 }}}
\right)\nonumber
\end{eqnarray}\mbox{} \\
where
\begin{equation}
\label{eq51}
\begin{array}{l}
 F_1 (y)=\int {\exp (-\frac{y^2}{W^2})\cos (\frac{k_g y^2}{2R})dy,} \\
 F_2 (y)=\int {\exp (-\frac{y^2}{W^2})\sin (\frac{k_g y^2}{2R})dy,} \\
 \end{array}
\end{equation}\mbox{} \\
are the quasi-probability integrals.
\begin{figure}[htbp]
\centerline{\includegraphics[scale=1.0]{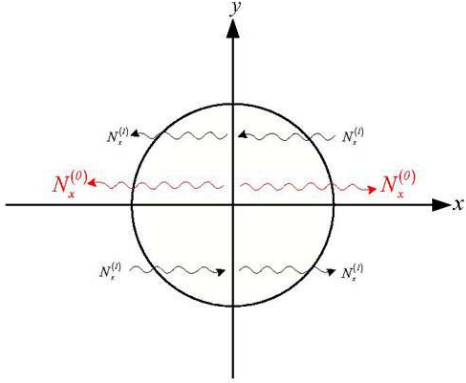}} \label{fig1}
\caption{ $N_x ^{(1)}$(signal) and $N_x ^{(0)}$(background) in 1st
(x,y,z$>$0), 2nd (x$<$0, y,z$>$0), 3rd (x,y$<$0,z$>$0) and 4th
(x$>$0,y$<$0,z$>$0) octants. $N_x ^{(1)}$and $N_x ^{(0)}$propagate
along opposite directions in the regions of 1st and 3rd octants,
while they have the same propagating directions in the region of 2nd
and 4th octants.}
\end{figure}

\begin{figure}[!h]
\centerline{\includegraphics[scale=1.0]{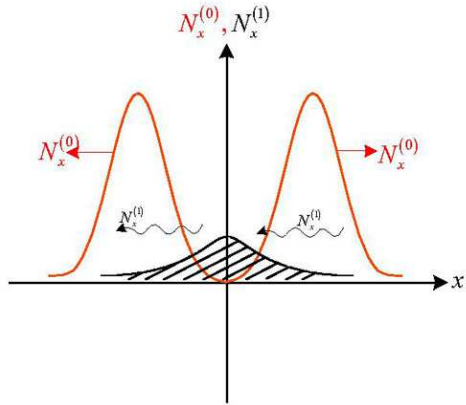}} \label{fig2}
\caption{Schematic diagram of strength distribution of $N_x ^{(1)}$
and $N_x ^{(0)}$in 1st and 2nd octants. We take note of that $\left|
{N_x ^{(0)}} \right|_{x=0} =\mbox{0}$ while $\left| {N_x ^{(1)}}
\right|_{x=0} =\left| {N_x ^{(1)}} \right|_{\max } $, and $N_x
^{(0)}$ is ``the outgoing wave'' to the \textit{yz-}plane.}
\end{figure}\clearpage

\begin{figure}[htbp]
\centerline{\includegraphics[scale=1.0]{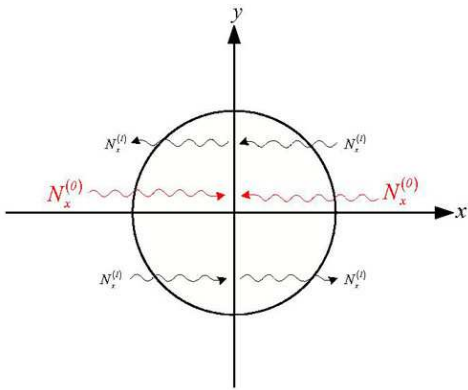}} \label{fig3}
\caption{$N_x ^{(1)}$ and $N_x ^{(0)}$ in the 5th (x,y$>$0,z$<$0),
6th (x$<$0, y$>$0, z$<$0), 7th (x,y,z$<$0) and 8th (x$>$0,y,z$>$0)
octants. $N_x ^{(1)}$and $N_x ^{(0)}$propagate along opposite
directions in the regions of 6th and 8th octants, while they have
the same propagating directions in the regions of 5th and 7th
octants.}
\end{figure}

\begin{figure}[htbp]
\centerline{\includegraphics[scale=0.95]{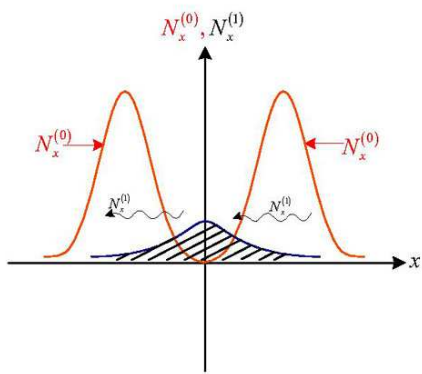}} \label{fig4}
\caption{Schematic diagram of strength distribution of $N_x ^{(1)}$
and $N_x ^{(0)}$ in the 5th and 6th octants. Also, we take note of
that $\left| {N_x ^{(0)}} \right|_{x=0} =0$ while $\left| {N_x
^{(1)}} \right|_{x=0} =\left| {N_x ^{(1)}} \right|_{\max .} $.
Unlike Fig.2, here $N_x ^{(0)}$ is ``the imploding wave'' to the
\textit{yz}-plane.}
\end{figure}
It is very interesting to compare $n_x^{(0)} $, Eq. (\ref{eq25}), and $n_x^{(1)} $,
Eq. (\ref{eq50}). From Eqs. (\ref{eq22}) and (\ref{eq25}), we can see that $\tilde {E}_y^{(0)} =0$
at the surface $x$=0, thus $n_x^{(0)} \vert _{x=0} =0$; while numerical
calculation shows that $n_x^{(1)} \vert _{x=0} $ has maximum. This means
that any photon measured by a detector (a high-sensitivity microwave
receiver) from $n_x^{(1)} \vert _{x=0} $ will be a signal of the EM
perturbation produced by the GW. Nevertheless, in the regions of $x\ne 0$,
we have $n_x^{(0)} \ne 0$. At first sight $n_x^{(1)} $ will be swamped by
the background $n_x^{(0)} $, so that $n_x^{(1)} $ has no observable effect
in this region. However, it will be shown that $n_x^{(1)} $ and $n_x^{(0)} $
propagate along the opposite directions in some local regions, and they have
the different rates of decay. Thus $n_x^{(1)} $ and $n_x^{(0)} $ can be
separated by the special fractal membranes (see below), so that $n_x^{(1)} $
(signal), in principle, would be observable. The total PPF passing through a
certain ``typical receiving surface'' $\Delta s$ at the \textit{yz}-plane will be\mbox{} \\
\begin{equation}
\label{eq52}
N_x^{(1)} =\int\!\!\!\int\limits_{\Delta s} {n_x^{(1)} \vert _{x=0} dydz} .
\end{equation}\mbox{} \\
Notice that $N_x^{(1)} $ is a unique non-vanishing photon flux
passing through the surface i.e., a number of photons per second.
Eqs. (\ref{eq50}) and (\ref{eq51}) show that $n_x^{(1)} $ is an even
function of the coordinates $x$, thus $n_x^{(1)} $ has the same
propagating direction in the regions of $x>0$ and $x<0$; and at the
same time, $n_x^{(1)} $ is an odd function of the coordinate y, so
the propagating directions of $n_x^{(1)} $ are anti-symmetric to the
regions of $y>0$ and $y<0$(such property ensured conservation of the
total momentum in the coherent resonance interaction). Considering
the outgoing (and imploding, i.e., they go in both directions)
property of $N_x^{(0)} $ in the region $z>0$(and $z<0)$ (this is a
typical property of the GB \cite{55}), it can be seen that
$N_x^{(1)}$ and $N_x^{(0)} $ propagate along opposite directions in
the regions of 1st ($x,y,z>0)$, 3rd ($x,y<0$,$z>0)$, 6th
($x<0,y>0,z<0)$ and 8th ($x>0$,$y,z<0)$ octants of the reacting
region between the magnetic poles, while they have the same
propagating directions in the regions of 2nd, 4th, 5th and 7th
octants. (see FIG1,FIG2,FIG3 and FIG4). In our EM system example,
all of the following parameters are chosen to exhibit values that
can be realized in the proposed laboratory experiments that is, they
are state of the art: \mbox{} \\

(1) P=10W, the power of the GB. In this case, $\psi _0 \approx 1.26\times
10^3Vm^{-1}$ for the GB of the spot radius $W_0 =0.05m$.

(2) $\hat {B}_y^{(0)} =3T$, the strength of the background static magnetic
field.

(3) $0\le y\le W_0 $, $0\le z\le 0.3\mbox{m}$, the integration
region $\Delta s$ (the receiving surface of the PPF) in Eq.
(\ref{eq52}), i.e., $\Delta s\approx 10^{-2} m^2$.

(\ref{eq1}) $l=l_2 +l_1 =0.3m$ and 6m, the interacting dimensions (or reacting
region) between the relic GW and the static magnetic field.

(\ref{eq2}) $\nu _e =\nu _g =5GHz(\lambda _g =0.06m)$, this is
typical frequency of the HFRGWs in the microwave band
\cite{1,2,3,4,5,6}, and of the HFGW predicted by possible
high-energy laboratory schemes \cite{9,10}.
\begin{figure}[htbp]
\centerline{\includegraphics[scale=1.0]{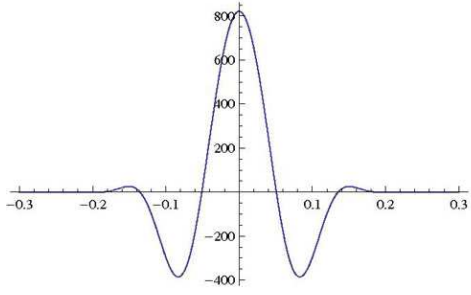}} \label{fig5}
\caption{The perturbative photon flux $N_x^{(1)} \left( {s^{-1}}
\right)$ generated by the HFGW of $h_{rms}  = 10^{ - {\rm{30}}}
/\sqrt {Hz}$ and $\nu =5GHz$, here detecting bandwidths $\Delta \nu
=1Hz$, $\left| {N_x^{(1)} } \right|=\left| {N_x^{(1)} }
\right|_{\max } =8.21\times 10^2\mbox{ s}^{\mbox{-1}}$ at $x=0$, we
take note of that the background photon flux $\left. {N_x^{(0)} }
\right|_{x=0} =0$[see, Eqs.(\ref{eq22}) and (\ref{eq25})], thus
$N_x^{(1)} $ would be an observable value, and $N_x^{(1)} $ and
$N_x^{(0)} $ propagate along opposite directions in the first
octant. }
\end{figure}
\begin{figure}[htbp]
\label{fig6} \centerline{\includegraphics[scale=1.0]{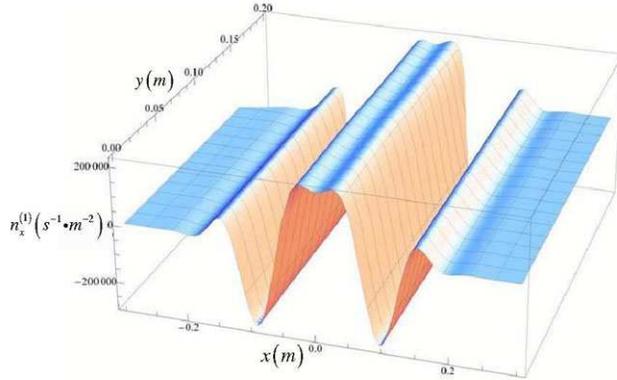}}
\caption{Two-dimensional distribution of the perturbative photon
flux density $n_x^{(1)} \left( {s^{-1}m^{-2}} \right)$ Eq.
[\ref{eq50}],where $z=l_2 $=0.3m, $l_1 =5.7m$, i.e., $z+l_1 =l_2
+l_1 =6m$,\mbox{ }0$<$y$<$0.2m,\mbox{ }$A_ \otimes = 10^{ - 30}
/\sqrt {Hz}$.It is shown that $\left| {n_x^{(1)} } \right|$has
maximum distribution in the region of -3.5cm$<$x $<$3.5cm.}
\end{figure}
\begin{table}[htbp]
\caption{The $ x$-component of PPFs and relevant parameters. Here A
is the root-means square value of the HFGW amplitudes, $l$ is
interacting dimensions between the HFGWs and the static magnetic
field, $N_x^{\left( 0 \right)} $and $N_x^{\left( 1 \right)} $ are
the $ x$-components of BPF and PPF, respectively.}
\begin{center}
\begin{tabular}{cccccc}
\hline $A\left( {Hz} \right)^{-\frac{1}{2}}$ &~~~ $l=l_1 +l_2 (m)$ &
~~~$N_x^{\left( 0 \right)} (s^{-1})$ &~~$N_x^{\left( 1 \right)}
(s^{-1})$ &~~~ $N_x^{\left( 0 \right)} (s^{-1})$ &~~
$N_x^{\left( 1 \right)} \left( {s^{-1}} \right)$  \\
 & & \multicolumn{2}{c}{$x=0\left( {cm} \right)$} &
\multicolumn{2}{c}{$x=3.5\left( {cm} \right)$}  \\
 \hline$10^{-22}$& 6& 0&~ $8.21\times 10^{10}$&$1.24\times
 10^{22}$&
$3.54\times 10^{10}$ \\
 $10^{-24}$& 6& 0& $8.21\times 10^8$& $1.24\times 10^{22}$&
$3.54\times 10^8$ \\
 $10^{-26}$& 6& 0& $8.21\times 10^6$& $1.24\times 10^{22}$&
$3.54\times 10^6$ \\
 $10^{-28}$& 6& 0& $8.21\times 10^4$& $1.24\times 10^{22}$&
$3.54\times 10^4$ \\
 $10^{-30}$& 6& 0& $8.21\times 10^2$& $1.24\times 10^{22}$&
$3.54\times 10^2$ \\
 $10^{-32}$& 6& 0& $8.21$& $1.24\times 10^{22}$&
$3.54$ \\
$10^{-34}$& 6& 0& 0& $1.24\times 10^{22}$&
0 \\
\hline
\end{tabular}
\label{tab1}
\end{center}
\end{table} \clearpage

FIG 5 gives result of numerical calculation for $N_x^{(1)} $.FIG 6
is two dimensional description of the numerical calculation for the
perturbative photo flux density $n_x^{(1)} $, Eq. (\ref{eq50}). From
Eqs.(\ref{eq50}), (\ref{eq51}) and (\ref{eq52}), $N_x^{(0)} $and
$N_x^{(1)} $we obtained in a \textit{1Hz} bandwidth are listed in
\mbox{TABLE \ref {tab1}}.

\begin{figure}[htbp]
\centerline{\includegraphics[scale=0.9]{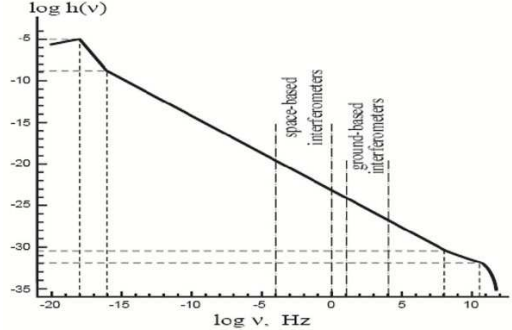}} \label{fig7}
\caption{Envelope of the $h_{rms} \left( \nu \right)$ spectrum for a
certain parameter condition. The figure is taken from
Ref.\cite{40}[P. Grishchuk, gr-qc/0504018]. The envelop shows that
the r.m.s. values of the HFRGW amplitudes in the region of
10$^{8}$-10$^{10}$Hz would be $\sim 10^{ - 30}  - 10^{ - 32} /\sqrt
{Hz} $,roughly.}
\end{figure}
In fact, the expected root-mean-square (rms) value $h_{rms} $ in the
GHz band of the dimensionless amplitudes by the different
cosmological models and parameters are quite
different\cite{1,2,3,4,5,6,42,51}. According to optimistic
estimation, their orders may be $h_{rms} $$\sim$${10^{-29}-10^{-30}}
\mathord{\left/ {\vphantom {{10^{-29}-10^{-30}} {\sqrt {Hz} }}}
\right. \kern-\nulldelimiterspace} {\sqrt {Hz} }$, while
conservative estimation may be only $h_{rms}
$$\sim$${10^{-34}-10^{-35}} \mathord{\left/ {\vphantom
{{10^{-34}-10^{-35}} {\sqrt {Hz} }}} \right.
\kern-\nulldelimiterspace} {\sqrt {Hz} }$. Ref. [40] provides a more
average estimation for the r.m.s value $h_{rms}
$$\sim$${10^{-30}-10^{-32}} \mathord{\left/ {\vphantom
{{10^{-30}-10^{-32}} {\sqrt {Hz} }}} \right.
\kern-\nulldelimiterspace} {\sqrt {Hz} }$ (see FIG 7). Thus in order
to detect the HFRGWs in the GHz band, the minimal detectable
amplitudes of the detecting systems would be $h $ $\sim$ $10^{-30}$
or less at least. Moreover, one often estimates the amplitudes of
relic GWs by their energy spectra, this is useful because it allows
us to quickly evaluate the cosmological importance of the generated
field in a given frequency interval. However, as pointed out by
Grishchuk \cite{39}, the primary and more universal concept is the
amplitude, not the spectrum density. It is the field, not its energy
density, which is directly measured by the GW detector. Therefore,
we listed the PPFs under the different amplitude conditions
($h_{srm} $$\sim$${10^{-22}-10^{-34}} \mathord{\left/ {\vphantom
{{10^{-22}-10^{-34}} {\sqrt {Hz} }}} \right.
\kern-\nulldelimiterspace} {\sqrt {Hz} })$ in a \textit{1Hz}
bandwidth in TABLE \ref {tab1}. Of course, possible distribution
region of the amplitude magnitudes of the HFRGWs may be only $_{
}h_{rms} $$\sim$${10^{-30}-10^{-32}} \mathord{\left/ {\vphantom
{{10^{-30}-10^{-32}} {\sqrt {Hz} }}} \right.
\kern-\nulldelimiterspace} {\sqrt {Hz} }$, there are no so strong
HFRGWs of $h_{rms} $$\sim$${10^{-22}-10^{-28}} \mathord{\left/
{\vphantom {{10^{-22}-10^{-28}} {\sqrt {Hz} }}} \right.
\kern-\nulldelimiterspace} {\sqrt {Hz} }$, but the estimation of the
PPFs can display detecting ability and sensitivity of the EM system
in the different amplitude conditions and in the frequency region.

TABLE \ref{tab1} shows that the most interesting region would be ``typical
receiving surface $\Delta s$'' at the \textit{yz}-plane (i.e., plane
of $x=0)$, where $\left. {N_x^{(0)} } \right|_{x=0} =0$ while
$\left. {N_x^{(1)} } \right|_{x=0} $has a maximum (e.g., if
$A=h_{rms} ={10^{-30}} \mathord{\left/ {\vphantom
{{\mbox{10}^{\mbox{-30}}} {\sqrt {Hz} }}} \right.
\kern-\nulldelimiterspace} {\sqrt {Hz} }$ and $l=6m$, then $\left.
{N_x^{(1)} } \right|_{x=0} =8.21\times 10^2s^{-1})$.

We emphasize that for the HFRGW and for the constant amplitude plane
HFGW, even if they have the same amplitude $h_{rms} ={10^{-30}}
\mathord{\left/ {\vphantom {{10^{-30}} {\sqrt {Hz} }}} \right.
\kern-\nulldelimiterspace} {\sqrt {Hz} }$ and the frequency $\nu
=5GHz$, their perturbative effects will be different. For the
constant amplitude plane HFGW propagating along the symmetrical axis
$z$ of the GB, it corresponds to a graviton flux of $N_g =3.77\times
10^{16}s^{-1}$ at the cross section of the waist of the GB (here the
minimum spot radius of the GB is equal to 5 cm). Unlike the constant
amplitude plane HFGW, due to the random property of the HFRGWs, they
contain every possible propagating direction, thus, as mentioned
above (Section \uppercase\expandafter{\romannumeral2}), the
propagating directions of the relic gravitons are nearing of state
of isotropy. In this case, only small fraction of the relic
gravitons will pass through the cross section of the GB. However,
the PPF generated by the resonant coherence modulation in our EM
system is the first-order perturbation rather than the second-order
perturbation of usual cavity EM response to HFGWs. Therefore, the
strength of PPF is proportional to the square root $\sqrt {N_g } $
of the graviton flux [i.e., it is proportional to the amplitude of
the GW, see Eq. (\ref{eq50})] and not the graviton flux itself $N_g
$ (i.e., the amplitude squared of the GW). In this case, numerical
calculation shows that if the deviation angle from the z-axis of the
propagating direction of the relic graviton flux is less than 10
degrees, then its perturbative effect and that of the graviton flux
propagating along the positive direction of the z-axis (i.e., best
resonant direction, see Section
\uppercase\expandafter{\romannumeral5}) are nearly the same.
Consequently, if all relic gravitons propagating along the deviation
angle region ($\theta \le 10^0)$ and passing through the cross
section of the GB are included, then the relic graviton flux at the
cross section will be $N_g \approx 2.89\times 10^{14}s^{-1}$ at
least. This means that in this case the gap between the PPFs
produced by the HFRGW and the constant amplitude HFGW will be about
1-2 orders of magnitude: this is satisfactory. Notice that then
ratio of the square roots of such
graviton fluxes will be \mbox{}\\
\begin{equation}
\label{eq53}
\sqrt {\frac{N_{g\mbox{ }relic\mbox{ }GW} }{N_{g\mbox{ }plane\mbox{ }GW} }}
=\sqrt {\frac{2.89\times 10^{14}}{3.77\times 10^{16}}} \approx 8.76\times
10^{-2}.
\end{equation}\mbox{}\\
From Eqs. (\ref{eq21})-(\ref{eq23}), (\ref{eq32}) and (\ref{eq42}),
we can calculate the 02-component  $\mathop {T^{02} }\limits^{(1)} $
of the first-order perturbation, and the corresponding PPF will be
$c \mathord{\left/ {\vphantom {c {\hbar \omega _e }}} \right.
\kern-\nulldelimiterspace} {\hbar \omega _e }\mathop
{T^{02}}\limits^{(1)} $, it expresses the first-order PPF density
$n_y^{(1)} $ propagating along the y-direction. By using the similar
means, we get $n_y^{(1)} $as follows:

\begin{eqnarray}
\label{eq54}
  n_y^{(1)}&=&\frac{c}{{\hbar \omega _e }} < \mathop {T^{02} }\limits^{(1)}  > _{\omega _e  = \omega _g }  =  - \frac{c}{{\mu _0 \hbar \omega _e }} < F_\alpha ^{0(0)} \tilde F^{2\alpha (1)}  + \tilde F_\alpha ^{0(1)} F^{2\alpha (0)}  > _{\omega _e  = \omega _g } \nonumber \\
  &=&-\frac{1}{{\hbar \omega _e }} < \frac{1}{{\mu _0 }}\tilde E_x^{(1)} \tilde B_z^{(0)}  > _{\omega _e  = \omega _g }  =  - \frac{1}{{2\mu _0 \hbar \omega _e }}{\mathop{\rm Re}\nolimits} \left\{ {\tilde E_x^{(1) * } \left( {\frac{{\partial \psi _x }}{{\partial y}} - \frac{{\partial \psi _y }}{{\partial x}}} \right)} \right\}_{\omega _e  = \omega _g }\nonumber \\
 &=& \frac{1}{{\hbar \omega _e }}\left\{ {\frac{{A_ \oplus  \hat B_y^{(0)} \psi _0 k_g y\left( {z + l_1 } \right)}}{{4\mu _0 \left[ {1 + {z \mathord{\left/
 {\vphantom {z f}} \right.
 \kern-\nulldelimiterspace} f}} \right]^{\frac{1}{2}} \left( {z + {{f^2 } \mathord{\left/
 {\vphantom {{f^2 } z}} \right.
 \kern-\nulldelimiterspace} z}} \right)}}\cos \left( {\frac{{k_g r^2 }}{{2R}} - \tan ^{ - 1} \frac{z}{f}} \right)} \right. - \frac{{A_ \oplus  \hat B_y^{(0)} \psi _0 y\left( {z + l_1 } \right)}}{{2\mu _0 W_0^2 \left[ {1 + ({z \mathord{\left/
 {\vphantom {z f}} \right.
 \kern-\nulldelimiterspace} f})^2 } \right]^{\frac{3}{2}} }} \nonumber\\
 &&\left. {\sin \left( {\frac{{k_g r^2 }}{{2R}} - \tan ^{ - 1} \frac{z}{f}} \right)} \right\}\exp \left( { - \frac{{r^2 }}{{W^2 }}} \right) + \frac{1}{{\hbar \omega _e }}\left\{ {\left( {1 - \frac{{4x^2 }}{{W^2 }}} \right)\frac{{A_ \oplus  \hat B_y^{(0)} k_g \left( {z + l_1 } \right)}}{{4\mu _0 R[1 + \left( {{z \mathord{\left/
 {\vphantom {z f}} \right.
 \kern-\nulldelimiterspace} f}} \right)^2 ]^{\frac{1}{2}} }}} \right. \nonumber\\
 &&\left[ {F_1 (y)\cos \left( {\frac{{k_g x^2 }}{{2R}} - \tan ^{ - 1} \frac{z}{f}} \right) - F_2 (y)\sin \left( {\frac{{k_g x^2 }}{{2R}} - \tan ^{ - 1} \frac{z}{f}} \right)} \right] - \\
 &&\left[ {\frac{2}{{W^2 }} + \left( {\frac{{k_g^2 }}{{R^2 }} - \frac{4}{{W^4 }}} \right)x^2 } \right]\frac{{A_ \oplus  \hat B_y^{(0)} \psi _0 \left( {z + l_1 } \right)}}{{4\mu _0 \left[ {1 + ({z \mathord{\left/
 {\vphantom {z f}} \right.
 \kern-\nulldelimiterspace} f})^2 } \right]^{\frac{1}{2}} }}\left[ {F_1 (y)\sin \left( {\frac{{k_g x^2 }}{{2R}}} \right.} \right. \nonumber\\
 &&\left. {\left. {\left. { - \tan ^{ - 1} \frac{z}{f}} \right) + F_2 (y)\cos \left( {\frac{{k_g x^2 }}{{2R}} - \tan ^{ - 1} \frac{z}{f}} \right)} \right]} \right\}\exp \left( { - \frac{{x^2 }}{{W^2 }}} \right).\nonumber
\end{eqnarray}
By comparing with Eqs. (\ref{eq50}) and (\ref{eq54}), we can see
that (1) $n_y^{(1)} $ is also an even function of the coordinates x
and an odd function of the coordinates y. Thus $n_y^{(1)} $ has the
same propagating direction in the regions of $x>$0 and x$<$0, and
$n_y^{(1)} $has opposite propagating direction in the regions of
y$>$0 and y$<$0. However, unlike property of $\left. {n_x^{(1)} }
\right|_{x=0} =n_{x\max }^{(1)} $, $\left. {n_y^{(0)} }
\right|_{y=0} =0$ and we also have $\left. {n_y^{(1)} }
\right|_{y=0} =0$. Therefore, $n_y^{(1)} $ and $n_y^{(0)} $ have
very similar distribution and behavior. In other words, in almost
all regions, $n_y^{(1)} $ will be swamped by the background
$n_y^{(0)} $, i.e., $n_y^{(1)} $ has no observable effect. (2) In
our case, $n_x^{(1)} $depends only on the state of $\otimes
$polarization of the HFRGW and it is independent of the state of
$\oplus $polarization of the HFRGW [see, Eq. (\ref{eq50})], while
$n_y^{(1)} $ depends only on the state of $\oplus $polarization
state and is independents of the state of $\otimes $polarization
[see, Eq. (\ref{eq54})]. Thus, the state of polarization displayable
in the EM system will be only the $\otimes $polarization component
of the HFRGW rather than the $\oplus $polarization component.

The quantum picture of the above-mentioned process can be described
as the resonant interaction of the photons with the gravitons in a
background of virtual photons (the statistic magnetic field) as a
catalyst \cite{14, 62}, i.e., the interaction involving elastic
scattering of the photons by the gravitons in the background of
virtual photons (in the reacting region between the magnet poles),
which can greatly increase the interaction cross section between the
photons and the gravitons. In other words, the interaction may
effectively change the physical behavior (e.g., propagating
direction, distribution, polarization, and phase) of the partial
photons in the local regions, and it does not require the resonant
conversion of the gravitons to the photons, the latter corresponds
to an extremely small conversion rate \cite{11}. Consequently, even
if the net increase of the photon number (the EM energy) of the
entire EM system approaches zero, then one still might find an
observable effect. In this case the requirements of relative
parameters can be greatly relaxed, such properties may be very
useful in order to detect the very weak signal of the HFRGWs. In the
case of astrophysical phenomenon, an analogous example is deflection
of light (an EM wave beam) in a gravitational field, which causes
the deflection of the propagating direction of the light ray, and
although there is no any change of the photon number, there is an
observable effect. Of course in this process the interacting
gravitational fields are static (e.g., the gravitational field of
the Sun). Thus there is no the frequency resonant effect between the
GWs and the EM waves and the space accumulation effect caused by the
coherent interaction of the two kinds of waves in the propagating
direction, but huge celestial gravitational fields compensate for
such a shortcoming. In our system the change of the propagating
directions and distribution of the partial photons in the local
regions is caused by the GW, while the strong background static
magnetic field provides a catalyst to enhance the resonant effect
between the EM wave (the photon flux) and the GW (gravitons), whose
coupling compensates in part for the weakness of the HFRGWs.

\section{The selection of the perturbative photon fluxes}
\begin{figure}[htbp]
\centerline{\includegraphics[scale=1.0]{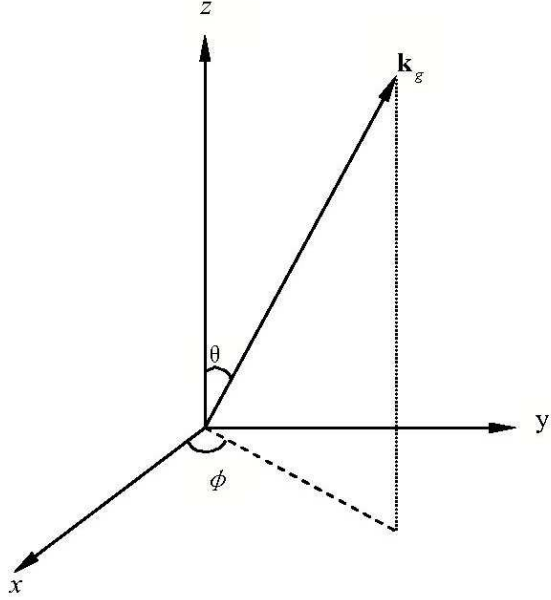}} \label{fig8}
\caption{The $z$-axis is the symmetrical axis of the Gaussian beam,
$\mbox{k}_g $ represents the propagating direction of the arbitrary
component of the relic GW.}
\end{figure}

 Because of the random property of the relic GWs, detection of the
relic GWs will be more difficult than that of the constant amplitude
plane GWs. However, we shall show that only the relic GW component
propagating along the positive direction of the $z$-axis can
generate optimal resonant response. It is true that for the relic GW
components propagating along the $ x-, y-$ axes and negative
direction of the $z$-axis, even if $\omega _g =\omega _e $, the PPFs
produced by them will be much less than that generated by the relic
GW component propagating along the positive direction of the
$z$-axis. Thus the perturbations produced by the relic GW components
propagating along the different directions cannot be counteracted.
In FIG 8 we draw the symmetrical axis (the $z$-axis) of the Gaussian
beam and the propagating directions $k_g $ of the arbitrary
component of the relic GWs.

In order to compare the PPFs generated by the different components of the
HFRGW, we shall discuss the perturbations caused by the HFRGW's components
propagating along some typical directions.

\subsection{The PPFs generated by the HFRGW components propagating
along different directions}

Here we assume $A=h_{rms} ={10^{-30}} \mathord{\left/ {\vphantom {{10^{-30}}
{\sqrt {Hz} }}} \right. \kern-\nulldelimiterspace} {\sqrt {Hz} },\mbox{ }\nu
_e =\nu _g =5GHz$, the detecting bandwidth is one Hz.

(a) $\theta =0$, i.e., the HFRGW component propagates along the
positive direction of the $z$-axis. As is calculated: the PPF
generated by the component may reach up to $8.21\times 10^2s^{-1}$
in a surface of $10^{-2}m^2$ area, (see, TABLE \ref{tab1}).

(b) $\theta =\pi $, i.e., the HFRGW component propagates along the negative
direction of the $z$-axis.

By using the similar means, one finds\mbox{}\\
\begin{eqnarray}
\label{eq55}
n_x^{(1)}&=&-\frac{1}{{\hbar \omega _e }} \cdot \left\{ {\frac{{A_ \otimes  \hat B_y^{(0)} \psi _0 k_g y(l_2  - z)}}{{4\mu _0 [1 + (z/f)^2 ]^{1/2} (z + f^2 /z)}}} \right.\sin \left( {2k_g z + \frac{{k_g r^2 }}{{2R}} - \tan ^{ - 1} \frac{z}{f}} \right)\nonumber \\
 &&\left.{+\frac{{A_ \otimes  \hat B_y^{(0)} \psi _0 y(l_2  - z)}}{{2\mu _0 W_0^2 [1 + (z/f)^2 ]^{3/2} }}\cos \left( {2k_g z + \frac{{k_g r^2 }}{{2R}} - \tan ^{ - 1} \frac{z}{f}} \right)} \right\} \cdot \exp \left( { - \frac{{r^2 }}{{W^2 }}} \right) \nonumber\\
 &&- \frac{1}{{\hbar \omega _e }}\left\{ {\left( {1 - \frac{{4x^2 }}{{W^2 }}} \right)\frac{{A_ \otimes  \hat B_y^{(0)} \psi _0 k_g (l_2  - z)}}{{4\mu _0 R\left[ {1 + (z/f)^2 } \right]^{1/2} }}} \right.\left[ {F_1 (y)\sin \left( {2k_g z + \frac{{k_g x^2 }}{{2R}} - \tan ^{ - 1} \frac{z}{f}} \right)} \right.\nonumber \\
 &&\left. { + F_2 (y)\cos \left( {2k_g z + \frac{{k_g x^2 }}{{2R}} - \tan ^{ - 1} \frac{z}{f}} \right)} \right] + \left[ {\frac{2}{{W^2 }} + \left( {\frac{{k_g ^2 }}{{R^2 }} - \frac{1}{{W^4 }}} \right)x^2 } \right] \\
 &&\frac{{A_ \otimes  \hat B_y^{(0)} \psi _0 (l_2  - z)}}{{4\mu _0 [1 + (z/f)^2 ]^{1/2} }}\left[ {F_1 (y)\cos \left( {2k_g z + \frac{{k_g x^2 }}{{2R}} - \tan ^{ - 1} \frac{z}{f}} \right)} \right. \nonumber\\
 &&- F_2 (y)\left. {\left. {\sin \left( {2k_g z + \frac{{k_g x^2 }}{{2R}} - \tan ^{ - 1} \frac{z}{f}} \right)} \right]} \right\}\exp \left( { - \frac{{x^2 }}{{W^2 }}} \right).\nonumber
\end{eqnarray}\mbox{}\\
\begin{table}[htbp]
\caption{The PPFs generated by the resonant HFGW components
propagating along the different directions, here $\hat
{B}^{(0)}=3T$,$A_\otimes,A_\oplus\sim10^{-30}/\sqrt{Hz}$, $\nu _g
=5GHz$, $l_2 +l_1 =6m$,}\mbox{} \\
\begin{center}
\begin{tabular}{p{140pt}cc}
\hline Propagating directions of the
 resonant components of the
relic HFGWs&~~~~~~~~~~~~~~~~~~~
$N_x^{(1)} $ (s$^{-1})$ ~~~\\
\hline ~~~~~~~~z &~~~~~~~~~~~~~~~~~~~$8.21\times 10^2$~~~\\
 ~~~~~~~~-z &~~~~~~~~~~~~~~~~~~~2.04$\times $10~~~\\
  ~~~~~~~~x &~~~~~~~~~~~~~~~~~~~4.07$\times 10^{-1}$~~~\\
   ~~~~~~~~y&~~~~~~~~~~~~~~~~~~~0 ~~~\\
\hline
\end{tabular}
\label{tab2}
\end{center}
\end{table}
Different from Eq. (\ref{eq50}), each and all terms in Eq.
(\ref{eq55}) contain oscillating factor $2k_g z$. We emphasize that
$2kz\approx 209z$ for the high-frequency relic GW of $\nu _g =5GHz$,
the factor $2kz$ will play a major role in the region of the
effective coherent resonance. In other words, the sign of $n_x^{(1)}
$ will oscillate quickly and quasi-periodically change as the
coordinate $z$ in the region increases. Thus the total effective PPF
passing through a certain ``typical receiving surface'' will be much
less than that generated by the relic GW component propagating along
the positive direction of the $z$-axis, (see Eq. (\ref{eq50}) and
TABLE \ref{tab2}).

(c) $\theta =\pi /2,\mbox{ }\phi \mbox{=0}$, i.e., the propagating direction
of the relic GW component is not only perpendicular to the symmetrical
z-axis of the GB, but also perpendicular to the static magnetic field $\hat
{B}_y^{(0)} $ directed along the y-axis so that it is along the x-axis. Here
we assume that the dimension of the $x$-direction of $\hat {B}_y^{(0)} $ is
localized in the region $-l_3 \le x\le l_4 $. Utilizing the similar means
the first-order perturbative EM fields generated by the direct interaction
of the relic GW with the static magnetic field can be given by\mbox{}\\
\begin{eqnarray}
\label{eq56} \tilde {E}_y^{(1)}&=&\frac{i}{2}A_\oplus \hat
{B}_y^{(0)} k_g c(x+l_3 )\exp [i(k_g x-\omega _g
t)]+\frac{1}{4}A_\oplus \hat {B}_y^{(0)} c\exp [i(k_g x+\omega _g
t)],\nonumber\\
\tilde {B}_z^{(1)}&=&\frac{i}{2}A_\oplus \hat {B}_y^{(0)} k_g (x+l_3
)\exp [i(k_g x-\omega _g t)]-\frac{1}{4}A_\oplus \hat {B}_y^{(0)}
\exp [i(k_g x+\omega _g t)],\nonumber\\
\tilde {E}_z^{(1)} &=&-\frac{1}{2}A_\otimes \hat {B}_y^{(0)} k_g
c(x+l_3 )\exp [i(k_g x-\omega _g t)]+\frac{i}{4}A_\otimes \hat
{B}_y^{(0)} c\exp [i(k_g x+\omega _g t)],\\
\tilde {B}_y^{(1)} &=&\frac{1}{2}A_\otimes \hat {B}_y^{(0)} k_g
(x+l_3 )\exp [i(k_g x-\omega _g t)]+\frac{i}{4}A_\otimes \hat
{B}_y^{(0)} \exp [i(k_g x+\omega _g t)],\nonumber
 \end{eqnarray}
 \[\left( {-l_3 \le x\le l_4 } \right)\]\mbox{}\\
In this case the coherent synchro-resonance ($\omega _e =\omega _g
)$ between the perturbative fields, Eq. (\ref{eq56}), and the GB can
be expressed as the following of PPF density, i.e.,\mbox{}\\
\begin{eqnarray}
\label{eq57}
n_x^{(1)}&=&\frac{1}{\mu _0 \hbar \omega _e }\left[
{\langle \tilde {E}_y^{(1)} \tilde {B}_z^{(0)} \rangle +\langle
\tilde {E}_y^{(0)} \tilde {B}_z^{(1)} \rangle -\langle \tilde
{E}_z^{(1)} \tilde {B}_y^{(0)} \rangle }
\right]_{\omega _e =\omega _g } \nonumber\\
&=&\frac{1}{2\mu _0 \hbar \omega _e }Re\left\{ {\tilde
{E}_y^{(1)\ast } \left[ {\frac{i}{\omega _e }\left( {\frac{\partial
\psi _x }{\partial y}-\frac{\partial \psi _y }{\partial x}} \right)}
\right]+\psi _y^\ast \tilde {B}_z^{(1)} +\tilde {E}_z^{(1)\ast }
\left( {\frac{i}{\omega _e }\frac{\partial \psi _x }{\partial z}}
\right)} \right\}_{\omega _e =\omega _g } ,
\end{eqnarray}\mbox{}\\
where $\tilde {B}_y^{(0)} $ and $\tilde {B}_z^{(0)} $ are the $y-$
and $z-$ components of the magnetic filed of the GB, respectively,
the angular brackets denote the average over time. Notice that we
choose the GB of the transverse electric modes, so $\tilde
{E}_z^{(0)} =0$. By using the same method, we can calculate
$n_x^{(1)} $, Eq. (\ref{eq56}). For example, first term in Eq.
(\ref{eq57}) can be written as\mbox{}\\
\begin{eqnarray}
\label{eq58}
&&\frac{1}{{2\mu _0 \hbar \omega _e }}{\mathop{\rm Re}\nolimits}\left\{ {\tilde E_y^{(1)*} \left[ {\frac{i}{{\omega _e }}\left( {\frac{{\partial \psi _x }}{{\partial y}} - \frac{{\partial \psi _y }}{{\partial x}}} \right)} \right]} \right\}_{\omega _e  = \omega _g }\nonumber\\
&=&- \frac{1}{{\hbar \omega _e }}\left\{ {\frac{{A_ \oplus  \hat B_y^{(0)} \psi _0 k_g y(x + l_3 )}}{{4\mu _0 \left[ {1 + (z/f)^2 } \right]^{1/2} (z + f^2 /z)}}\sin \left [ {k_g (x - z) + \frac{{k_g r^2 }}{{2R}} - \tan ^{ - 1} \frac{z}{f}} \right]} \right. \nonumber\\
&&+ \left. {\frac{{A_ \oplus  \hat B_y^{(0)} \psi _0 y(x + l_3 )}}{{2\mu _0 W_0^2 \left[ {1 + (z/f)^2 } \right]^{3/2} }}\cos \left[ {k_g (x - z) + \frac{{k_g r^2 }}{{2R}} - \tan ^{ - 1} \frac{z}{f}} \right]} \right\}\exp ( - \frac{{r^2 }}{{W^2 }}) \nonumber\\
&&-\frac{1}{{\hbar \omega _e }}\left\{ {(1 - \frac{{4x^2 }}{{W^2 }})} \right.\frac{{A_ \oplus  \hat B_y^{(0)} \psi _0 k_g y(x + l_3 )}}{{4\mu _0 R\left[ {1 + (z/f)^2 } \right]^{1/2} }}\left[ {F_1 (y)\sin \left( {k_g (x - z) + \frac{{k_g x^2 }}{{2R}} - \tan ^{ - 1} \frac{z}{f}} \right)} \right] \nonumber\\
&&+ F_2 (y)\cos \left( {k_g (x - z) + \frac{{k_g x^2 }}{{2R}} - \tan ^{ - 1} \frac{z}{f}} \right) + \left[ {\frac{2}{{W^2 }} + (\frac{{k_g^2 }}{{R^2 }} - \frac{1}{{W^4 }})x^2 } \right] \\
&&\frac{{A_ \oplus  \hat B_y^{(0)} \psi _0 y(x + l_3 )}}{{4\mu _0 \left[ {1 + (z/f)^2 } \right]^{1/2} }}\left[ {F_1 (y)\cos \left( {k_g (x - z) + \frac{{k_g x^2 }}{{2R}} - \tan ^{ - 1} \frac{z}{f}} \right)} \right. \nonumber\\
&&\left. {\left. { - F_2 (y)\sin \left( {k_g (x - z) + \frac{{k_g
x^2 }}{{2R}} - \tan ^{ - 1} \frac{z}{f}} \right)} \right]}
\right\}\exp ( - \frac{{x^2 }}{{W^2 }}),\nonumber \mbox{}\\
\nonumber &&\quad \qquad\quad \qquad\quad \qquad\quad \qquad(- l_3
< x < l_4 )\nonumber
\end{eqnarray}\mbox{}\\
It can be shown that calculation for the 2nd and 3rd terms in Eq.
(\ref{eq57}) is quite similar to first term, and they have the same
orders of magnitude, we shall not repeat it here. Notice that unlike
$n_x^{(1)} $ produced by the relic GW component propagating along
the positive direction of the $z$-axis, the phase functions in Eq.
(\ref{eq58}) contain oscillating factor $k_g (x-z)$, and because it
is always possible to choose $l_2 +l_1 \gg l_4 +l_3 $, i.e., the
dimension of the $z$-direction of $\hat {B}_y^{(0)} $ is much larger
than its $x$-direction dimension. Thus, the PPF expressed by Eq.
(\ref{eq50}) will be much larger than that represented by Eq.
(\ref{eq57}) (see, TABLE \ref{tab2}).

(d) $\theta =\pi /2,\mbox{ }\phi =\pi /2$, i.e., the relic GW component
propagates along the y-axis, which is parallel with the static magnetic
field $\hat {B}_y^{(0)} $.

According to the Einstein-Maxwell equations of the weak field, then
the perturbation of the GW to the static magnetic field vanishes
[14, 28], i.e.,
\begin{equation}
\label{eq59}
n_x^{(1)} =0.
\end{equation}
It is very interesting to compare $n_x^{(1)} $ in Eqs. (\ref{eq50}), (\ref{eq55}), (\ref{eq58}) and
(\ref{eq59}), as is shown that although they all represent the PPFs propagating
along the x-axis, their physical behaviors are quite different. In the case
of $\theta =\phi =\pi /2$, $n_x^{(1)} =0$, Eq. (\ref{eq59}); when $\theta =\pi $ and
$\theta =\pi /2,\mbox{ }\phi =0$, the PPFs contain the oscillating factors
$2k_g z$ and $k_g (x-z)$, respectively [see Eqs. (\ref{eq55}) and (\ref{eq58})]. Only under
the condition $\theta =0$, does the PPF, Eq. (\ref{eq50}), not contain any
oscillating factor, but only a slow variation function in the z direction.
This means that $n_x^{(1)} $ produced by the relic GW component propagating
along the positive direction of the $z$-axis, has the best space accumulation
effect (see TABLE \ref{tab2}). Thus, as previously mentioned, our EM system would be
very sensitive to the propagating directions of the relic GWs. In other
words the EM system has a strong selection capability to the resonant
components from the stochastic relic GW background. Therefore, if real relic
GW background has a small deviation to the isotropy of space, then it should
be possible to provide an HFRGW map of the celestial sphere by changing the
direction of the symmetrical axis of the GB, or alternatively by the
utilization of multiple EM detectors.

\subsection{The separation of the PPFs (signal) from the BPFs}

\begin{figure}[htbp]
\centerline{\includegraphics[scale=1.0]{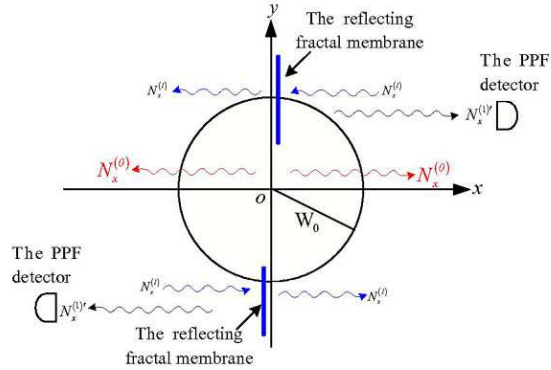}} \label{fig9}
\caption{$N_x^{(0)} $,$\mbox{ }N_x^{(1)} $ and $N_x^{(1)'}$ in the
1st and 3rd octants. After $\mbox{ }N_x^{(1)} $ is reflected by the
fractal membrane, (e.g., $N_x^{(1)'}$ in the 1st and 3rd octants),
$N_x^{(1)'}$ and $N_x^{(0)} $ will have the same propagating
direction. However, $\mbox{ }N_x^{(1)'}$ can keep its strength
invariant within one meter to the membrane (see, e.g., Refs.
\cite{56,57}), while $\mbox{ }N_x^{(0)} $ decays as the typical
off-axis (radial distance r) Gaussian decay rate $\exp (-2r^2/W^2)$
[see, Eq.(\ref{eq25})] and attenuated further by superconducting
baffles, then the ratio $N_x^{(1)'}/N_x^{(0)} $ would be larger than
one in the whole region of $0.35m<x<1m$, although $N_x^{(0)} \gg
N_x^{(1) '}$ in the region of $0<x<0.35m$.}
\end{figure}
\begin{figure}[htbp]
\centerline{\includegraphics[scale=1.0]{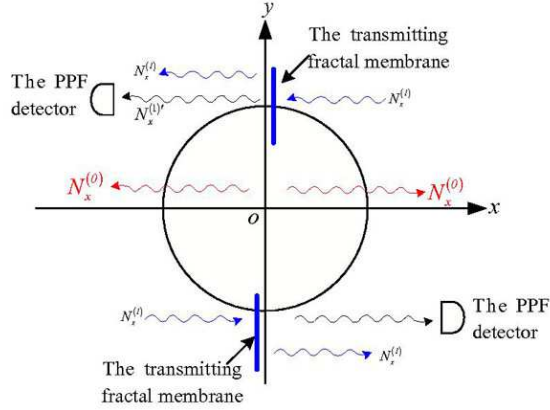}} \label{fig10}
\caption{$N_x^{(0)} $,$\mbox{ }N_x^{(1)} $ and $N_x^{(1)'}$ in the
2nd and 4th octants. Unlike Fig.9, here $N_x^{(1)'}$ is the PPF
transmitted by the transmitting fractal membrane, then the PPF
detectors should be put in the 2nd and 4th octants}
\end{figure}

\begin{table}[htbp]
\caption{Comparison of the PPF reflected or transmitted by the
fractal membrane and the BPF in the $x$-direction, here $\hat
{B}^{(0)}=3T$, $h_{rms}\sim10^{ - 30} /\sqrt {Hz} $, $\nu _g =5GHz$,
$l_2 +l_1 =6m$ and detecting bandwidth $\Delta \nu =1Hz$. The PPF
$N_x^{(1)} $ reflected or transmitted (defined as $N_{x}^{(1)'}) $by
the fractal membrane can nearly keep its strength invariant nearly
within one meter distance from the membrane \cite{56,57,58} (or even
more attenuated by superconductor baffles along the x-axis). Even if
according to most conservative estimation to the fractal membranes
\cite{68}, the photon flux reflected or transmitted by the fractal
membranes can keep ninety percent of its strength at the position of
one meter distance from the fractal membranes. Here our comparison
is just from this conservative estimation. Thus the $N_{x}^{(1)'}$
and $N_x^{(0)} $ would have a comparable order of magnitude in the
region $35$cm$<x<37$cm (over a diffraction-limited spot area of
$\sim $3$\times $10$^{-4}$ m$^{2})$.} \mbox{} \\\mbox{} \\
\begin{center}
\begin{tabular}{p{140pt}ccccc}
\hline The distance to the fractal membrane(cm)& 0& 3.50& 32.59&
35.09&
37.00 \\
\hline$N_x^{(0)} $(s$^{-1})$ & 0& 1.24$\times $10$^{22}$&
6.73$\times $10$^{5}$& 8.20$\times $10$^{2}$&
3.50 \\
\hline$N_{x}^{(1)'}$(s$^{-1})$ ~~~&8.21$\times $10$^{2}$~~~&
8.18$\times $10$^{2}$~~~& 7.94$\times $10$^{2}$~~~& 7.92$\times
$10$^{2}$~~~&
7.90$\times $10$^{2}$ \\
\hline
\end{tabular}
\label{tab3}
\end{center}
\end{table}

In recent years new types of fractal membranes have been
successfully developed \cite{56,57,58}. Firstly, these fractal
membranes can provide nearly total reflection for the EM waves
(photon flux) with certain frequencies in the GHz band; at the same
time, they can provide nearly total transmission for the photon
fluxes with other frequencies in the GHz band (the fractal-membrane
pattern can be ``significantly sub-wavelength in all dimensions''
\cite{56}). Secondly, the photon fluxes reflected and transmitted by
the fractal membranes can keep their strength invariant within the
distance of one meter from the fractal-membrane's surface,
especially if the fractal-membrane reflectors are back-to-back
very-shallow (or segmented) paraboloid mirrors that focus the PPF on
the detectors situated out along opposite ends of the x-axis. In
this case the diffracted focus spot at each detector exhibits a
radius of $\lambda _{g}$/$\pi=6$cm/$\pi\sim $1.91cm (area of $\sim
10^{-4} m^{2})$ [9,10]. Thirdly, such frequencies can be regulated
in the GHz band. Since $N_{x}^{(1)}$ (signal) and $N_{x}^{(0)}$
(background) propagate along the negative and positive directions of
the x-axis in the first octant (the region of $x,y,z>0)$,
respectively, i.e., $N_{x}^{(1)}$ propagates along the direction
toward the fractal membrane, while $N_{x}^{(0)}$ propagates along
the direction away from the fractal membrane (see FIG 9). Using the
reflecting fractal membranes with their plane or paraboloid faces
normal to the x-axis, it will reflect only $N_{x}^{(1)}$ and not
$N_{x}^{(0)}$. Once $N_{x}^{(1)}$ is reflected (defined as
$N_{x}^{(1)'}$)it will have the same propagating direction as
$N_{x}^{(0)}$. However, after $N_{x}^{(1)}$ is reflected, it can
keep its strength invariant within one meter distance from the
fractal membrane [56, 57], while $N_{x}^{(0)}$ decays as the typical
Gaussian decay rate $\exp (-\frac{2r^2}{W^2})$[see, Eq.
(\ref{eq25})] to each side of the GB (x and y directed), then the
ratio $N_{x}^{(1)'}/N_{x}^{(0)}$ (the signal-to-background noise
ratio in the x-direction) would be larger than one in the whole
region of $0.35\mbox{m}\le x\le 1m$ (see, TABLE \ref{tab3}, $x$ is
the distance from the detectors to the fractal membranes). TABLE
\ref{tab3} shows that the BPF $N_{x}^{(0)}$ is much larger than the
PPF $N_{x}^{(1)'} $in the region $0<x<35cm$, while the $N_{x}^{(0)
}$and $N_{x}^{(1)'} $have the same order of magnitude at
$x=35.09cm$, and $N_{x}^{(1)'} $would be larger than $N_{x}^{(0)}$
in the region of $x>35cm$. In other words, in the region the
signal-to-background noise ratio ${N_x^{(1)'}} \mathord{\left/
{\vphantom {{N_x^{(1)'}} {N_x^{(0)} }}} \right.
\kern-\nulldelimiterspace} {N_x^{(0)} }$in the x-direction might
gain up to a comparable order of magnitude. It appears better to use
the transmitting fractal membranes, because the PPF transmitted by
the fractal membrane can also keep its strength invariant within one
meter to the membrane, and the PPF does not change its propagating
direction. In this case, the PPF detectors in 1st and 3rd octants in
FIG 9 should be replaced by the detectors in 2nd and 4th octants in
FIG 10.

In fact, the circular polarized ``monochromatic component''
,Eq.(35), is often called the right-handed circular polarization,
while a left-handed circular polarized component has following form
\begin{eqnarray}
&& h_\oplus =h_{11} =-h_{22} =A_\oplus \exp \left[ {i\left( {k_g
z-\omega_g t}\right)} \right], \nonumber\\
&& h_\otimes =h_{12} =h_{21} =-iA_\otimes \exp \left[ {i\left( {k_g
z-\omega_g t}\right)} \right],
\end{eqnarray}
where $A_\oplus , A_\otimes \approx A\left( {k_g } \right)/a\left( t
\right)$. In our EM system, according to Eqs.(35),(38),(39) and
(69), the propagating direction of $N_{x}^{\left( 1 \right)} $
depends on the choice of circular polarization. Thus, if the
interacting ``monochromatic component'' is only the left-handed
circular polarized state, Eq.(69), then the propagating direction of
$N_{x}^{\left( 1 \right)} $ will be opposite to that generated by
the right-handed circular polarized component, Eq.(35), and then
$N_{x}^{\left( 1 \right)} $ and $N_x^{\left( 0 \right)} $ propagate
along opposite directions in the regions of 2nd, 4th, 5th and 7th
octants, while they have the same propagating direction in the
regions of 1st, 3rd, 6th and 8th octants. In such a case, the
distinguishable PPF from the BPF would be $N_x^{\left( 1 \right)} $
in the regions of 2nd and 4th octants but not in the regions of 1st
and 3rd octants.

If the both circular polarizations exist at the same time (In this
case the two polarized states often have a certain phase difference.
More detailed investigation for the issues will be done
elsewhere),then the PPFs (here we defined them as $N_{xI}^{\left( 1
\right)} $ and $N_{xII}^{\left( 1 \right)} $, respectively)
generated by the right- and left-handed polarized circular
components will propagate along the opposite directions in the every
octant. One of them propagates along the positive direction in the
x-axis, and another one the negative direction in the x-axis.
However,because acting effects of fractal membranes to
$N_{xI}^{\left( 1 \right)} $ and $N_{xII}^{\left( 1 \right)} $ are
quite different, one kind of the two PPFs ($N_{xI}^{\left( 1
\right)} $ or $N_{xII}^{\left( 1 \right)} $) could be distinguished
from the BPF.

For example, in the first-octant (the region of x,y,z$>$0)
$N_{xI}^{\left( 1 \right)} $ and $N_{xII}^{\left( 1 \right)} $
propagate along the negative and positive directions in the x-axis,
respectively. This means that $N_{xI}^{(1)} $ propagates along the
direction toward the fractal membrane, while $N_{xII}^{(1)} $ and
$N_x^{(0)} $ propagate along the direction away from the fractal
membrane (see also, Figs. 1, 9, and 10). In this case the PPF
reflected (or transmitted) by the fractal membrane will be only
$N_{xI}^{(1)} $ but not $N_{xII}^{(1)} $ and $N_x^{(0)} $. Once
$N_{xI}^{(1)} $ is reflected (or transmitted) by the fractal
membrane, it will keep its strength invariant within one meter
distance from the fractal membrane [56, 57], while $N_x^{(0)} $
decays as the typical Gaussian decay rate exp($-\frac{2r^2}{W^2})$,
$N_{xII}^{(1)} $ decay as the exp($-\frac{x^2}{W^2})$ [see also,
Eq.(59)]. Therefore, the ratio $N_{xI}^{(1)} /N_x^{(0)} $ would has
a comparable order of magnitude in the distance of 35cm$<$x$<$37cm
from the fractal membrane (see also, Table III). In this case, in
principle, $N_{xI}^{(1)} $ can still be distinguished from
$N_x^{(0)} $, while the other one of the PPFs $N_{xII}^{(1)} $ will
be swamped by $N_x^{(0)} $ due to the same propagating direction and
the similar decay way of them. There is a similar property in the
second octant (the region of x$<$0, y, z$>$0), unique difference is
that where $N_{xII}^{(1)} $ and $N_x^{(0)} $ propagate along the
opposite directions, while $N_{xI}^{(1)} $ and $N_x^{(0)} $
propagate along the same direction. Thus the distinguishable PPF
from the BPF would be only $N_{xII}^{(1)} $ but not $N_{xI}^{(1)} $.
Utilizing the similar means it can be shown that in the 1st, 2nd,
3rd and 4th octants, the distinguishable PPF from the BPF will be
one kind of the PPFs, namely, $N_{xI}^{(1)} $ or $N_{xII}^{(1)} $ .
Consequently, role of the fractal membranes looks like a ``one-way
valve" with strong focusing function to the photon flux in the GHz
band. This property will be very useful to distinction and
displaying the PPFs generated by the stochastic HFRGW background.

Of course, if considering other possible noise sources and
diffraction effects, the values listed in TABLE \ref{tab1} will be
further reduced, thus an obvious gap still exists between the
theoretical schemes and reality.

\section{The thermal noise and the EM noise.}

At the moment there are no operating prototypes of the EM detecting system,
although relevant researches and construction of the EM detecting system are
already in progress, it is difficult to give a complete description for the
noise issues. However, since our purpose is display and detection of the PPF
of about $\nu $=5GHz in the terminal microwave receiver, our attention will
be focused into two key aspects:

(1). What are strength and physical behavior of the PPF (signal) and the BPF
(background) reaching the microwave receiver;

(2). How to distinguish the PPF and other photons caused by noise, such as
the thermal noise, background noise and external EM noise. Here we shall
give a very brief and rough review.

Except for the background photon noise issue just mentioned, there
are the thermal noise sources and possible external EM noise
sources. Because the frequency of the PPF (signal) is roughly
$5GHz$, if the system is cooled down to $KT<\hbar \omega _e $ ($K$
is Boltzmann's constant, $\omega _e =2\pi \nu _e ,\mbox{ }\nu _e
=5GHz)$, i.e., $T<\hbar \omega _e /K\sim 0.24K$, then the frequency
$\nu _m $ of the thermal photons will be less than the $\nu _e $ of
the PPF. If the apparatus is kept to a lower temperature, e.g.,
\textit{T $<0.024K$ or $24 mK$} (this is well within the current
technology), then we have $\nu _m \approx 10^{-2}\nu _e $. Thus the
difference in the frequency band for such two kinds of photons would
be very great, i.e., the signal photon flux and the thermal photons
can be easily distinguished. In other words, practically speaking
there are no thermal photons at $\mbox{5GHz}$, and in this way the
thermal noise can be suppressed as long as the EM detector can
select the correct frequency. Note that the low temperature is very
convenient for the operation of the superconductors and the strong
static magnetic field.

For the possible external EM noise sources, using a Faraday cage or
shielding covers made from such fractal membranes \cite{56,57,58},
or from a tight mosaic of superconductor chips on the inside surface
of the detector's cryogenic containment vessel, would be very
effective. Moreover, a good ``microwave darkroom'' can provide an
effective shielding environment, and in this case possible
dielectric dissipation (using a vacuum operation) can be effectively
suppressed. In this case one would obtain a suitable environment for
a measurable signal-to-noise ratio.

Also, the superposition of the relic GW stochastic components will cause the
fluctuation of the PPFs, even if such ``monochromatic components'' all
satisfy the frequency resonant condition ($\omega _e =\omega _g )$. However,
Eqs. (\ref{eq50}), (\ref{eq52}), (\ref{eq55}), (\ref{eq58}) and (\ref{eq59}) show that the metric perturbation only
influences the strength fluctuation of the PPFs and does not influence the
``direction resonance.'' That is, it does not influence the selection
capability of the EM system to the propagating directions of the relic GWs,
and it does not influence average effect over time of the PPFs.

In addition, the values of the PPFs discussed in the present paper
depend on the strength of the HFRGWs in the GHz band expected by the
QIM and other relevant string cosmology scenarios (e.g., see Refs,
\cite{1,2,3,4,5,6,7}). Because such models and scenarios are
somewhat controversial, we cannot know in advance how accurate these
models and scenarios might be. If the strength of the real HFRGWs in
the GHz band are much less than the magnitude expected by such
models and scenarios, even if the required conditions can be
satisfied and one might still not be able to detect and measure such
HFRGWs, then the HFRGW models will be corrected. Thus, this scheme
might provide an indirect way to test such models and scenarios,
that is, as suggested by Brustein et al. \cite{63}, a null
experiment would be valuable. In any event, the HFGW generator and
detector experiment described in Ref. \cite{9,10}, which operate at
about the same frequencies as the HFRGWs, will prove the concept of
the present detector independently from cosmological experiments.

Moreover, there are some issues and problems need further
investigation. For example, how to generate a typical and
high-quality GB, how to suppress distortion of the spot radius of
the GB and align it, what is concrete correction to the PPF caused
the higher order modes of the GB, how to further estimate and
analysis the relevant noise sources, what are concrete influence and
correction of the fractal membranes (or plates) to the GB itself,
how to estimate and effectively suppress diffraction effect by new
materials, (e.g. the fractal membranes), how to ensure a good vacuum
to avoid the scattering of photons and dielectric dissipation caused
by the dust and other particles, etc. All these issues and problems
need careful theoretical and experimental study. More detailed
investigation concerning such issues will be an object of further
research and will be studied elsewhere.

\section{Concluding Remarks}

\begin{enumerate}
\item Although usual analytic expressions of the relic GWs are often complicated, the high-frequency asymptotic behavior of them in the microwave band can be expressed as simpler forms, and they can be described as superposition of all quasi-monochromatic components. The energy density of the HFRGWs is positive definite, and their momentum densities have reasonable physical behavior, the EM resonant response of the HFRGWs in the laboratory frame of reference can be treated as resonance interaction of the quasi-monochromatic HFGWs with the EM fields.
\item Under the synchroresonance condition, coherent modulation of the HFRGW to the preexisting transverse BPFs would produce the transverse PPFs, the PPFs propagating along two orthogonal directions of the double transverse polarized electric modes of the GB are generated by the pure $\otimes $ polarization and the pure $\oplus $polarization states of the HFGW, respectively. The former has maximum at the longitudinal symmetrical surface of the GB where the transverse BPF vanishes, but the later and the BPF have the same distribution. Thus, the former may provide an observable effect while the latter will be swamped by the BPF.
\item The PPF reflected or transmitted by the fractal membranes exhibits a very small decay compared with the much stronger BPF. In our case this is the PPF produced by the pure $\otimes $ polarization state of the HFGW. Another interesting area would be the region in which the PPF and the BPF might reach up a comparable order of magnitude.
\item Although an obvious gap still exists between the theoretical estimation and reality, there are a potential advanced space and new ways [64-67] to further improve the sensitivity and the detecting ability of the EM system. These new ways and technology will include generation of super-strong static magnetic fields (e.g., use of crystal channel effect), ultra-high sensitivity microwave single photon detectors such as a circuit Quantum Electrodynamics device (CQED) photon detector, Ryberg Atom Cavity detector, SQUID array mux, Josephon Junction Arrays, etc., and possible optimized combination of them. They are possible to further narrow such gap and provide new promise.
\end{enumerate}

\begin{center}
\section*{Acknowledgements}
\end{center}
This work is supported by the National Basic Research Program of
China under Grant No.2003 CB 716300, the National Natural Science
Foundation of China under Grant 10575140, the Foundation of China
Academy of Engineering Physics under Grant No.2008 T0401,
No.2008T0402, the Nature Science Foundation of Chongqing under Grant
8562, GRAVWAVE {\textregistered} LLC, Transportation Sciences
Corporation and Seculine Consulting of the USA. \mbox{}

\end{document}